\documentclass[showpacs,amsmath,amssymb,twocolumn,pra,superscriptaddress]{revtex4-2}
\usepackage{amssymb}
\usepackage[dvips]{graphicx}
\usepackage{enumerate}
\usepackage{epsfig} 
\usepackage{subfigure}
\usepackage{xcolor}
\usepackage[T1]{fontenc}
\usepackage{fullpage}
\usepackage{amsthm,amsfonts,amssymb,amscd,mathrsfs,xspace,framed}
\usepackage{mathrsfs,amsmath}
\usepackage{color}
\usepackage{setspace}
\usepackage{url}
\usepackage{wrapfig}
\usepackage{tikz}
\usepackage{enumitem}
\usepackage{bm}
\usepackage{comment}
\usepackage{txfonts}
\usepackage{array}
\usepackage{color}
\usepackage{braket}
\usepackage{hyperref}
\usepackage[english]{babel}
\usepackage{url}
\usepackage{graphicx}
\usepackage[ruled,vlined]{algorithm2e}

\newcommand{\QT}[1]{{{\textcolor{black}{#1}}}}
\def\be{\begin{equation}}
\def\ee{\end{equation}}
\def\ba{\begin{eqnarray}}
\def\ea{\end{eqnarray}}

\makeatletter
\AtBeginDocument{\let\LS@rot\@undefined}
\makeatother

\def\be{\begin{equation}}
\def\ee{\end{equation}}
\def\ba{\begin{eqnarray}}
\def\ea{\end{eqnarray}}

\begin{document}

\title{Quantum-enhanced data classification with a variational entangled sensor network}

\author{Yi Xia}
\affiliation{
James C. Wyant College of Optical Sciences, University of Arizona, Tucson, Arizona 85721, USA
}

\author{Wei Li}
\affiliation{
Department of Materials Science and Engineering, University of Arizona, Tucson, Arizona 85721, USA
}

\author{Quntao Zhuang}
\affiliation{
Department of Electrical and Computer Engineering, University of Arizona, Tucson, Arizona 85721, USA
}
\affiliation{
James C. Wyant College of Optical Sciences, University of Arizona, Tucson, Arizona 85721, USA
}

\author{Zheshen Zhang}
\email{zsz@arizona.edu}
\affiliation{
Department of Materials Science and Engineering, University of Arizona, Tucson, Arizona 85721, USA
}
\affiliation{
James C. Wyant College of Optical Sciences, University of Arizona, Tucson, Arizona 85721, USA
}

\begin{abstract}
Variational quantum circuits (VQCs) built upon noisy intermediate-scale quantum (NISQ) hardware, in conjunction with classical processing, constitute a promising architecture for quantum simulations, classical optimization, and machine learning. However, the required VQC depth to demonstrate a quantum advantage over classical schemes is beyond the reach of available NISQ devices. Supervised learning assisted by an entangled sensor network (SLAEN) is a distinct paradigm that harnesses VQCs trained by classical machine-learning algorithms to tailor multipartite entanglement shared by sensors for solving practically useful data-processing problems. Here, we report the first experimental demonstration of SLAEN and show an entanglement-enabled reduction in the error probability for classification of multidimensional radio-frequency signals. Our work paves a new route for quantum-enhanced data processing and its applications in the NISQ era.

\end{abstract}

\maketitle
\section{Introduction}
The convergence of quantum information science and machine learning (ML) has endowed radically new capabilities for solving complex physical and data-processing problems~\cite{carleo2017solving,biamonte2017quantum,gao2018quantum,havlivcek2019supervised,schuld2019quantum,dunjko2018machine,rebentrost2014quantum,schutzhold2003pattern,moll2018quantum}. Many existing quantum ML schemes hinge on large-scale fault-tolerant quantum circuits composed of, e.g., quantum random access memories. At present, however, the available noisy intermediate-scale quantum (NISQ) devices~\cite{Preskill2018quantumcomputingin,wang2018multidimensional} hinder these quantum ML schemes to achieve an advantage over classical ML schemes. Recent developments in hybrid systems~\cite{zhu2019training,moll2018quantum} comprising classical processing and variational quantum circuits (VQCs) open an alternative avenue for quantum ML. In this regard, a variety of hybrid schemes have been proposed, including quantum approximate optimization~\cite{farhi2014quantum}, variational quantum eigensolvers~\cite{kandala2017hardware}, quantum multi-parameter estimation~\cite{meyer2020variational}, and quantum kernel estimators and variational quantum models~\cite{schuld2019quantum,havlivcek2019supervised}.
On the experimental front, hybrid schemes have been implemented to seek the ground state of quantum systems~\cite{kandala2017hardware,nam2020ground}, to perform data classification~\cite{havlivcek2019supervised}, to unsample a quantum circuit~\cite{carolan2020variational}, and to solve the MAXCUT problem~\cite{otterbach2017unsupervised,farhi2017quantum}. The finite quantum coherence time and circuit depths of state-of-the-art NISQ platforms, however, hold back a near-term quantum advantage over classical ML schemes. An imperative objective for quantum ML is to harness NISQ hardware to benefit practically useful applications~\cite{biamonte2017quantum}.
\begin{figure*}
    \centering
    \includegraphics[width=1\textwidth]{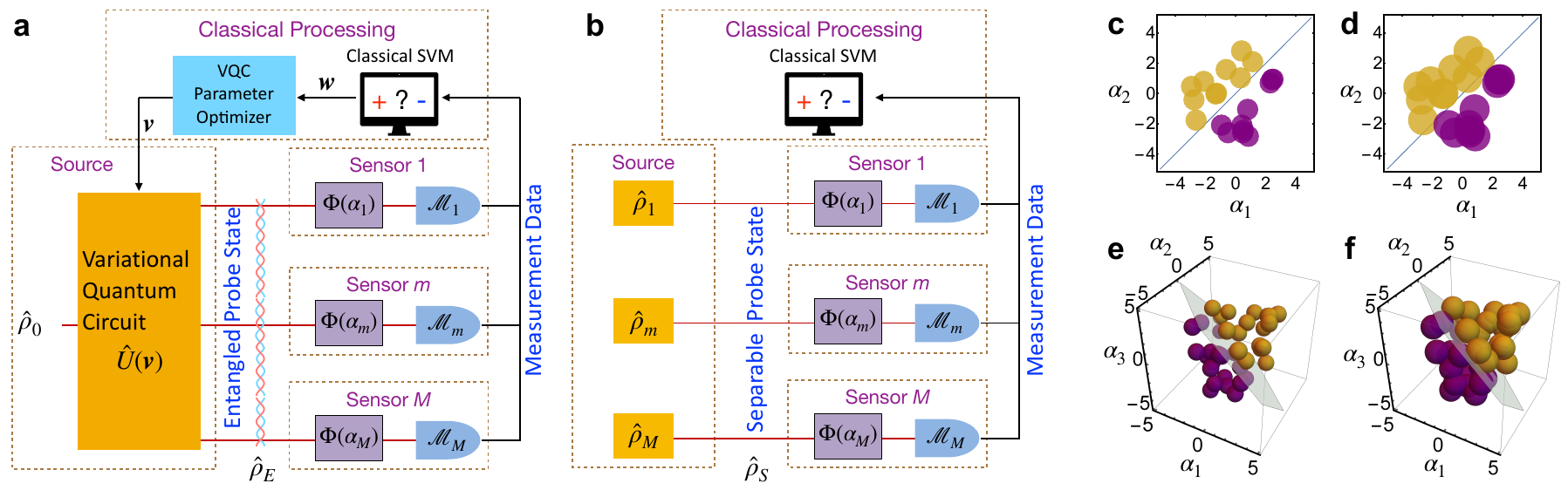}
    \caption{Schematics of SLAEN and classical classifier with sample data sets. (a) In SLAEN, a VQC is configured to generate an entangled probe state. In classical processing, measurement data are utilized to train a classical SVM, whose hyperplane $\bm w$ is mapped to the VQC setting $\bm v$ by the VQC parameter optimizer. (b) Classical classifier only uses a classical SVM. (c, d) 2D data acquired by two sensors, applicable to RF-field direction classification. (e, f) 3D data acquired by three sensors, applicable to RF-field mean-amplitude classification. Circle/sphere: data point with radius representing standard deviation of estimation uncertainty. Entangled sensors (c, e) with a clear error-probability reduction over classical separable sensors (d, f).}
    \label{fig:concept_class_data}
\end{figure*}

\section{Supervised learning assisted by an entangled sensor network}

A multitude of data-processing scenarios, such as classification of images captured by cameras~\cite{deng2009imagenet}, target detection through a phased array~\cite{fenn2000development}, and identification of molecules~\cite{baaske2012optical}, encompass sensors for data acquisition. Recent theoretical~\cite{zhuang2018distributed,ge2018distributed,proctor2018multi,qian2019heisenberg,gatto2019distributed,oh2020optimal,grace2020entanglement} and experimental~\cite{guo2020distributed,xia2020demonstration,liu2020distributed} advances in distributed quantum sensing have unleashed the potential for a network of entangled sensors to outperform classical separable sensors in capturing global features of an interrogated object. Such a capability endowed by distributed quantum sensing creates an opportunity to further utilize VQCs to configure the entangled probe state shared by the sensors to enable a quantum advantage in data-processing problems. 

Supervised learning assisted by an entangled sensor network (SLAEN)~\cite{zhuang2019physical} is such a hybrid quantum-classical framework empowered by entangled sensors configured by a classical support-vector machine (SVM) for quantum-enhanced high-dimensional data classification, as sketched in Fig.~\ref{fig:concept_class_data} (a). SLAEN employs a VQC parameterized by $\boldsymbol{v}$ to create an entangled probe state $\hat{\rho}_E$ shared by $M$ quantum sensors. The sensing attempt at the $m$th sensor is modeled by running the probe state through a quantum channel, $\Phi(\alpha_m)$, where the information about the object is embedded in the parameter $\alpha_m$. A measurement modeled by $\mathcal{M}_m$ on the output quantum state from the channel then yields $\tilde{\alpha}_m$ as the measurement data. To label the interrogated object, a classical SVM chooses a hyperplane parameterized by $\boldsymbol{w}$ to separate the measurement data into two classes in an $M$-dimensional space. To learn the optimum hyperplane and the configuration of the VQC that produces the optimum entangled probe state under a given classification task, the sensors first probe training objects with known labels, and the measurement data and the true labels are used to optimize the hyperplane $\boldsymbol{w}$ of the SVM. Then, the VQC parameter optimizer maps $\boldsymbol{w} \rightarrow \boldsymbol{v}$, which in turn configures the VQC to generate an entangled probe state $\hat{\rho}_E = \hat{U}(\boldsymbol{v})\hat{\rho}_0\hat{U}^\dag(\boldsymbol{v})$ that minimizes the measurement noise subject to the chosen hyperplane. As a comparison, Fig.~\ref{fig:concept_class_data} (b) sketches a conventional classical classifier that solely relies on a classical SVM trained by the measurement data obtained by separable sensors to seek the optimum hyperplane for classification. By virtue of the entanglement-enabled noise reduction, SLAEN yields a substantially lower error probability than that achieved by the classical classifier, which is illustrated and compared in Fig.~\ref{fig:concept_class_data} (c--f) for two classification problems in, respectively, a two-dimensional (2D) data space and a three-dimensional (3D) data space.

\begin{figure*}[t]
    \centering
    \includegraphics[width=1\textwidth]{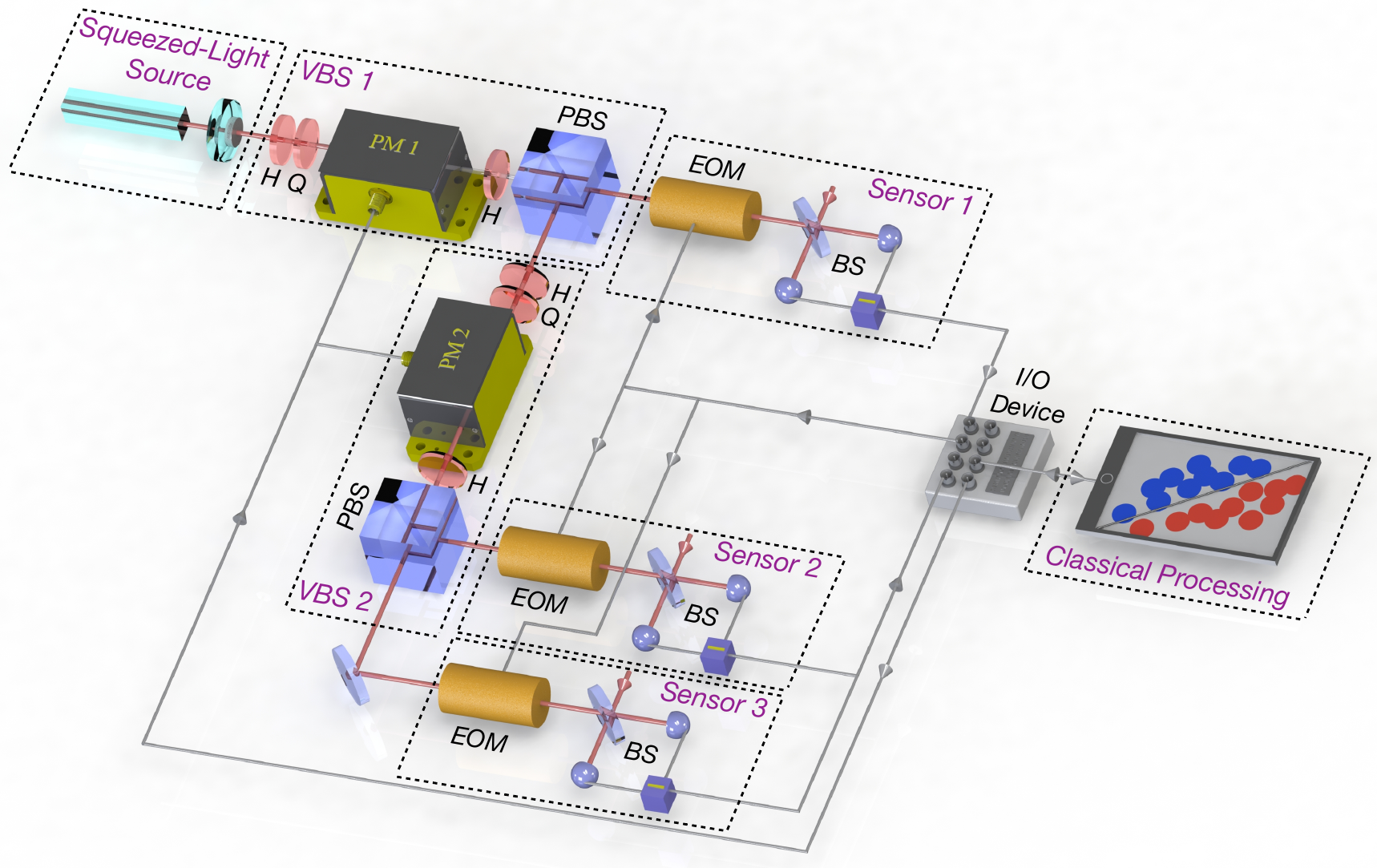}
    \caption{Experimental diagram. Squeezed light processed by two variable beam splitters (VBSs), each composed of two half-wave plates (H), a quarter-wave plate (Q), a phase modulator (PM), and a polarizing beam splitter (PBS), generating a three-partite entangled probe state. Each sensor comprises an electro-optic modulator (EOM) and a balanced homodyne measurement setup. Measurement data acquired by an I/O device and processed on a classical computer for training and data classification. During training, classical processing controls VBSs and EOMs through the I/O device. LO: local oscillator; BS: beam splitter.}
    \label{fig:exp}
\end{figure*}

\section{Experiment}
We demonstrate SLAEN in a quantum optics platform based on continuous-variable (CV) entanglement. The experiment endeavors to classify a feature embedded in a set of radio-frequency (RF) signals: $\left\{\mathcal{E}_m(t) = E_m \cos (\omega_c t + \varphi_m)\right\}_{m=1}^M$, where $\boldsymbol{E} \equiv \{E_m\}_{m=1}^M$ and $\boldsymbol{\varphi} \equiv \{\varphi_m\}_{m=1}^M$ are, respectively, the RF amplitudes and phases at the $M=3$ sensors, and $\omega_c$ is the RF carrier frequency. The class label $y$ is determined by a joint function of amplitudes and phases: $y = F(\boldsymbol{E},\boldsymbol{\varphi})$.

The experimental setup is sketched in Fig.~\ref{fig:exp}. An optical parametric amplifier source emits a single-mode squeezed state represented by the annihilation operator $\hat{b}$. To acquire data, a VQC prepares an entangled probe state, described by $\{\hat{b}_m\}_{m = 1}^M$, by applying a unitary operation $\hat{U}(\boldsymbol{v})$ on $\hat{b}$. The VQC setting is entailed in $\boldsymbol{v} \equiv \{v_m, \phi_m\}_{m=1}^M$, where $v_m$ is the power ratio of the squeezed state sent to the $m$th sensor, satisfying $\sum_{m=1}^M v_m = 1$, and $\phi_m$ is a phase shift imparted on the quantum state at the $m$th sensor. {The VQC is composed of two variable beam splitters (VBSs) and three phase shifters. A VBS comprises two half-wave plates (H), a quarter-wave plate (Q), a phase modulator (PM), and a polarizing beam splitter (PBS). The PM controls the splitting ratio of the VBS and thus determines $v_m$, while the phase shift $\phi_m$ is controlled by an RF signal delay (see Appendix B for details).} At the $m$th sensor, an electro-optic modulator (EOM) converts the RF signal into a displacement $\alpha_m \propto E_m \sin \varphi_m$ on the phase quadrature $\hat{p}_m \equiv (\hat{b}_m-\hat{b}_m^\dag)/2i$. Three homodyne detectors then measure the quadrature displacements, and the measurement data are diverted to a classical processing unit for training, classification, and VQC setting optimization.

SLAEN consists of a training stage and a utilization stage. The training stage is aimed at using $N$ training data points $\{\boldsymbol{E}^{(n)},\boldsymbol{\varphi}^{(n)},y^{(n)}\}_{n=1}^N$ supplied to the sensors to optimize the hyperplane used by the SVM and the entangled probe state. $y^{(n)}\in\{-1,+1\}$ is the true label for the $n$th training data point. The training data point leads to the homodyne measurement data $\tilde{\bm \alpha}^{(n)}$ from the sensors. $\tilde{\bm \alpha}^{(n)}$ and $y^{(n)}\in\{-1,+1\}$ are the only information available to the classical processing unit. For a hyperplane specified by $\left\{\bm w \equiv \{w_m\}_{m=1}^M, b\right\}$, we define a cost function
\begin{equation}
\mathcal{E}_\lambda(\bm w,b) = \sum_{n=1}^N \left\vert 1-y^{(n)} \left(\bm w \cdot \tilde{\bm \alpha}^{(n)}+b\right) \right\vert_++\lambda \|\bm w \|^2,
\label{eq:cost_function}
\end{equation}
where $\vert x\vert_+$ equals $x$ for $x\ge 0$ and zero otherwise, $\|\cdot\|$ is the usual two-norm, and $\lambda \|\bm w \|^2$ is used to avoid over-fitting. {The $\bm w \cdot \tilde{\bm \alpha}^{(n)}$ term represents a weighted average over the measurement data acquired by different sensors. It is the weighted average that benefits from using multipartite entanglement to reduce the measurement noise~\cite{xia2020demonstration,zhuang2019physical}.} Only the support vectors, i.e., points close to the hyperplane with $y^{(n)}\left(\bm w \cdot \tilde{\bm \alpha}^{(n)}+b\right)\le 1$, contribute non-trivially to the cost function. The rationale behind constructing such a cost function is that errors primarily occur on support vectors in a classification task, thus accounting for the deviations of all data points from the hyperplane in the cost function is non-ideal. 

To enable efficient minimization of the cost function, we adopt a stochastic optimization approach in which the hyperplane and the VQC setting are updated in each training step consuming a single data point. Suppose the optimized hyperplane is $\left\{\bm w^{(n-1)}, b^{(n-1)}\right\}$ after $\left(n-1\right)$ training steps. Prior to updating the hyperplane in the $n$th training step, the inferred label is derived by $\tilde{y}^{(n)}={\rm sign}\left(\bm w^{(n-1)} \cdot \tilde{\bm \alpha}^{(n)}+b^{(n-1)}\right)$. Using a simultaneous perturbation stochastic approximation (SPSA) algorithm, the hyperplane is updated to $\left\{\boldsymbol{w}^{(n)}, b^{(n)}\right\}$ (see Appendix A for algorithm details). Once an updated hyperplane is found, the VQC optimizer performs the mapping $\bm w^{(n)} \rightarrow \bm v^{(n)}$ to configure the VQC so that its generated entangled probe state minimizes the measurement noise subject to the current hyperplane. Specifically, one desires that the virtual mode $\hat{b}_v \equiv \sum_{m = 1}^M w_m^{(n)} \hat{b}_m$, whose phase-quadrature measurement outcome constitutes the $\bm w^{(n)}\cdot \tilde{\bm\alpha}^{(n+1)}$ term in $\tilde{y}^{(n+1)}$, is identical to the original squeezed-light mode $\hat{b}$ so that the overall uncertainty in labeling is minimized. This is accomplished by setting $\sqrt{v_m^{(n)}}\exp\left(i\phi_m^{(n)}\right) = w_m^{(n)}$ in the VQC parameter optimizer. Physically, this is the noise-reduction mechanism, stemming from the quantum correlations between the measurement noise at different sensors, that gives rise to SLAEN's quantum advantage over the classical classifier in which the measurement noise at different sensors is independently subject to the standard quantum limit. After $N$ training steps, the cost function is near its minimum with the hyperplane $\left\{\bm w^\star, b^\star\right\} \equiv \left\{\bm w^{(N)}, b^{(N)}\right\}$, and the VQC setting $\bm v^\star \equiv \bm v^{(N)} $. Then, in the utilization stage, SLAEN configures the VQC using $\bm v^\star$ and classifies the measurement data $\tilde{\bm \alpha}$ with an unknown label using the optimized hyperplane $\bm w^\star$:
\begin{equation}
\tilde{y}={\rm sign}\left(\bm w^\star \cdot \tilde{\bm \alpha}+b^\star\right).
\label{eq:label}
\end{equation}

\begin{figure*}[th]
    \centering
    \includegraphics[width=1\textwidth]{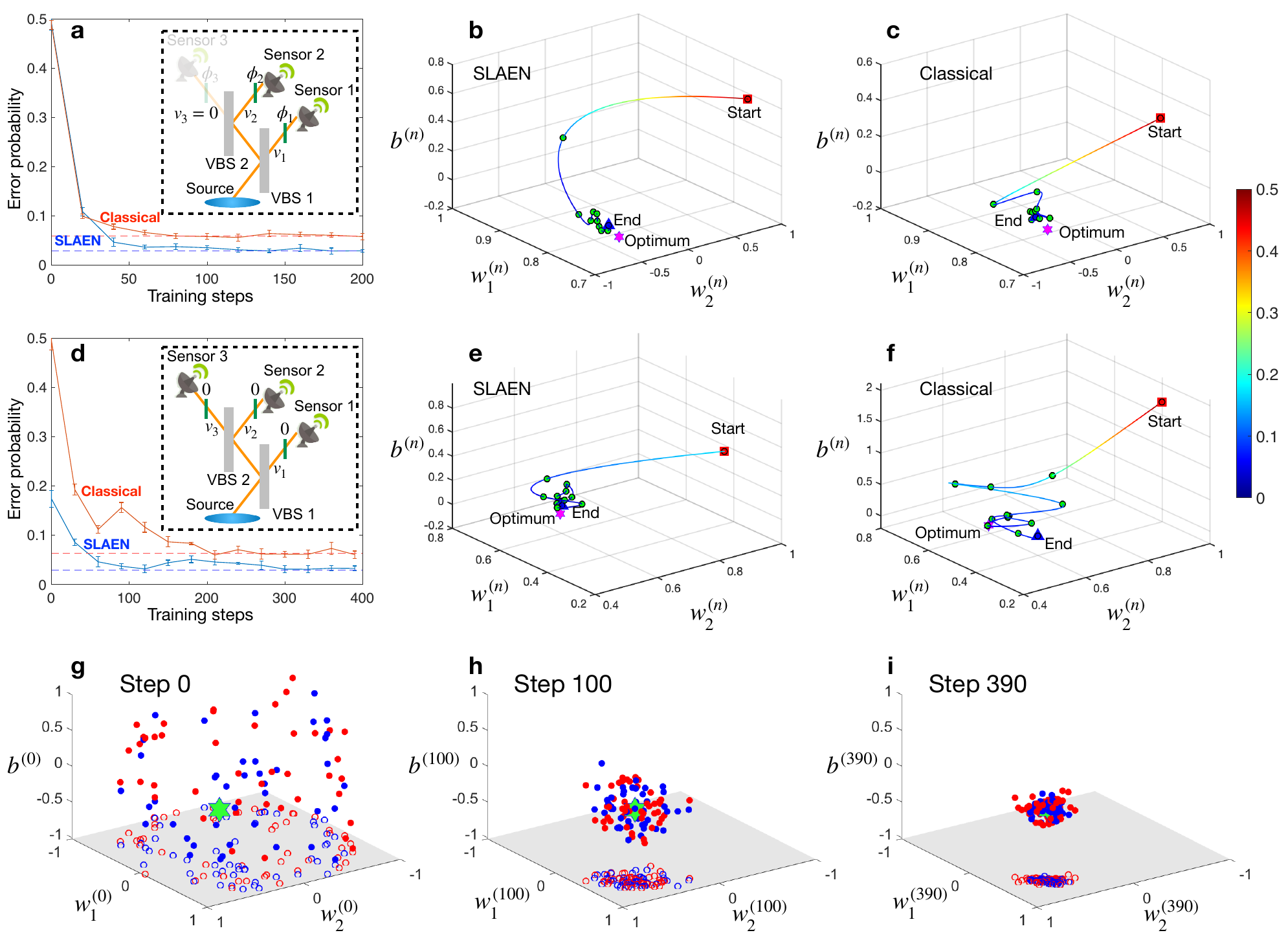}
    \caption{Experimental results for training SLAEN and classical classifier. Convergence of error probabilities during training for 2D data classification (a) and 3D data classification (d). Blue curves: SLAEN. Red curves: classical classifier. Horizontal dashed lines: expected error probabilities based on true hyperplanes and measurement-noise levels. Error bars: one standard deviation of uncertainty derived from five measurements each with 1000 data points. Insets: VQC parameters being optimized. VBS: variable beam splitter. History of hyperplane $\left\{\left(w_1^{(n)}, w_2^{(n)}\right), b^{(n)}\right\}$ during training for 2D data classification (b, c) and 3D data classification (e, f). (b, e) SLAEN; (c, f) classical classifier. Red squares: initial hyperplane parameters prior to training; blue triangles: hyperplane parameters after training; hexagrams: optimum hyperplane parameters. Color gradients: evolution of error probabilities during training. Green circles: samples of hyperplane parameters at every 20 (30) training steps for 2D (3D) data classification. Curves are obtained from a cubic spline data fitting. Simulated distributions of hyperplane parameters prior to training (g), at Step 100 (h), and at Step 390 (i). Blue filled circles: SLAEN hyperplanes; red filled circles: classical-classifier hyperplanes; hexagrams: optimum hyperplanes. Open circles: projected hyperplane parameters onto $(w_1,w_2)$ plane (grey). SLAEN's optimized hyperplanes distribute statistically closer to optimum solutions.}
    \label{fig:training}
\end{figure*}

SLAEN is a versatile framework capable of tailoring the entangled probe state and the classical SVM to enhance the performance of multidimensional data-classification tasks. In our experiment, SLAEN first copes with 2D data acquired by two entangled sensors, as illustrated in Fig.~\ref{fig:concept_class_data} (c). As an example and useful application for 2D data classification, we demonstrate the classification of the incident direction of an emulated RF field. To this end, the training commences with an initial hyperplane specified by $\left\{\bm w^{(0)} =\left(\sqrt{0.50}, \sqrt{0.50}\right), b^{(0)} = 0.70\right\}$, which is mapped to an initial VQC setting $\bm v_0 = \{0.50, 0.50, 0, 0, 0, 0\}$. The training stage comprises 200 steps each using a training data point with randomly generated RF-field phases and an associated label $\{{\bm \varphi^{(n)}},y^{(n)}\}^{200}_{n = 1}$, while the RF-field amplitudes are fixed equal at all sensors. Applying the training data $\bm \varphi^{(n)}$ on the EOMs at the two sensors leads to quadrature displacements $\bm \alpha^{(n)} = \{\alpha_1^{(n)}, \alpha_2^{(n)}\}$, whose each component is chosen to follow a uniform distribution in $[-4,4]$ (in the shot-noise unit). The signal-to-noise ratio of the data set is tuned by excluding the data points within a margin of $\epsilon$ from the hyperplane while the total number of training data points is fixed at 200. In doing so, the signal-to-noise ratio is raised as $\epsilon$ increases. The true labels for the RF-field directions is derived by the RF-phase gradient: $y^{(n)}= {\rm sign} \left(\varphi_1^{(n)}-\varphi_2^{(n)}\right) = {\rm sign} \left(\bm w_t \cdot \bm \alpha^{(n)}\right)$, where $\left\{\bm w_t = \left(\sqrt{1/2}, -\sqrt{1/2}\right), b_t = 0\right\}$ parameterize the true hyperplane. The true labels are disclosed while $\left\{\bm w_t, b_t\right\}$ and $\bm \alpha^{(n)}$ are kept unknown to SLAEN. The optimization for the SVM hyperplane and the VQC setting then follows.

As a performance benchmark, we train the classical classifier, using the identical training data in training SLAEN, to undertake the 2D data-classification task. Unlike SLAEN, the classical classifier uses a separable probe state $\hat{\rho}_S$ to acquire the measurement data, which are then used to train the classical SVM to seek a hyperplane that minimizes the classification error probability. In the experiment, the squeezed-light source is turned off while applying the same training data as those used for SLAEN, thereby ensuring an equitable performance comparison. The initial hyperplane prior to the training is randomly picked as $\left\{\bm w^{(0)} = (0.67, 0.74), b^{(0)} = 0.39\right\}$. In the absence of entanglement-enabled noise reduction, a higher error probability is anticipated for the classical classifier, as illustrated in Fig.~\ref{fig:concept_class_data} (d).

The effectiveness of the training for SLAEN and the classical classifier is demonstrated by the converging error probabilities measured at different training steps, as plotted in Fig.~\ref{fig:training} (a). The inset describes the VQC parameters being optimized. The convergence of the error probabilities beyond 100 training steps indicates that near-optimum settings for the hyperplanes and the VQC have been found. With such optimized parameters, SLAEN is able to generate an entangled probe state that minimizes the measurement noise, as illustrated in Fig.~\ref{fig:concept_class_data} (c) and compared to Fig.~\ref{fig:concept_class_data} (d) for the case of the classical classifier by a set of sample data points represented by the circles, whose radii correspond to the standard deviation of estimation uncertainty. 

SLAEN and the classical classifier are next trained to tackle 3D data-classification problems. As an example, we demonstrate the classification of the sign for the RF-field mean amplitude across three sensors. The training in either scenario uses 390 data points $\{\bm E^{(n)},y^{(n)}\}_{n = 1}^{390}$ with randomly generated RF-field amplitudes, while the RF phases are fixed at $\bm \varphi^{(n)} = 0$. The true labels are then given by $y^{(n)}= {\rm sign} \left(\sum_{m = 1}^3 E_m^{(n)}\right) = {\rm sign} \left(\bm w_t \cdot \bm \alpha^{(n)}\right)$, where $\left\{\bm w_t = \left(\sqrt{1/3}, \sqrt{1/3}, \sqrt{1/3}\right), b_t = 0\right\}$ specify the true hyperplane, which unknown to SLAEN and the classical classifier. The error probabilities during training for both scenarios are plotted in Fig.~\ref{fig:training} (d), with its inset describing the VQC parameters being optimized. The error probabilities converge after 250 training steps, indicating that near-optimum settings for the hyperplanes and the VQC have been found. Once both are trained, SLAEN shows a clear error-probability advantage over that of the classical classifier, as observed in Fig.~\ref{fig:training} (d) and intuitively illustrated in Fig.~\ref{fig:concept_class_data} (e) and (f). 

The trajectories of the evolving hyperplane $\left\{\bm w^{(n)}, b^{(n)}\right\}$ during training are plotted in Fig.~\ref{fig:training} (b, c) for 2D data classification and (e, f) for 3D data classification. The hexagrams entail the optimum hyperplane parameters. The hyperplane parameters approach the optimum with a decreasing error probability during training, as anticipated. Notably, the optimized hyperplanes obtained by SLAEN are considerably closer to the true hyperplanes, i.e., the optimum solutions, than those attained by the classical classifier thanks to SLAEN's reduced measurement noise. To further investigate SLAEN's improved accuracy to problem solutions, we randomly generate 50 sets of initial hyperplanes for SLAEN and the classical classifier and plot in Fig.~\ref{fig:training} (g--i) the simulated distributions of the hyperplanes at different steps of training for 3D data classification. The simulation shows that SLAEN's optimized hyperplanes (red circles) have a distance of $d_{S} = 0.135 \pm 0.056$ to the true hyperplane, i.e., the optimum solutions (hexagrams), as compared to a distance of  $d_{C} = 0.167 \pm 0.073$ for the optimized hyperplanes attained by the classical classifier (red circles) (see Appendix C for simulation results and comparison with experiment).

\begin{figure}[bth]
    \centering
    \includegraphics[width=0.45\textwidth]{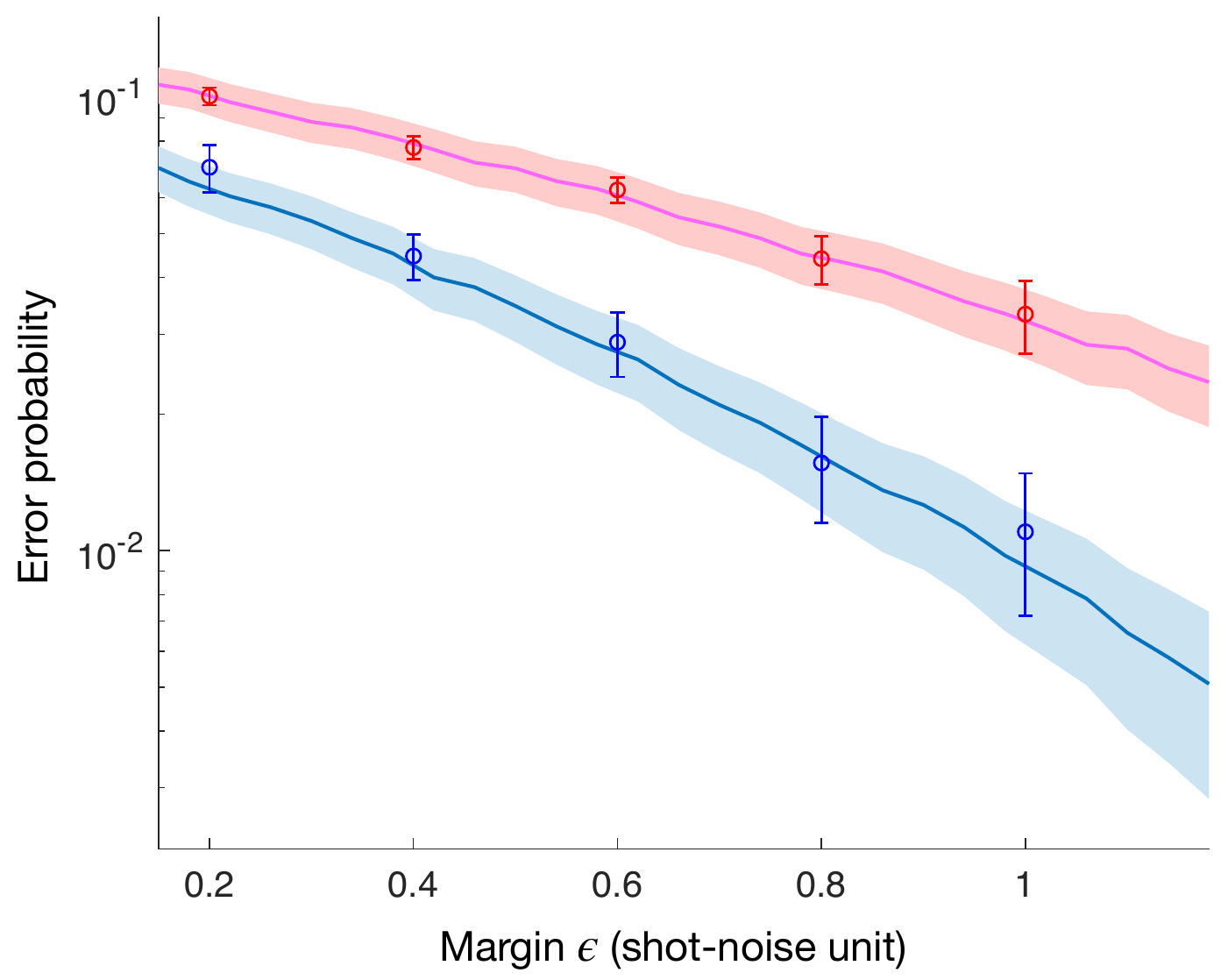}
    \caption{Scaling of error probability vs. margin of the data set. Blue: SLAEN; red: classical classifier. Circle: estimated error probability based on 5 sets of 1000 experimental data points. Solid lines: error probabilities obtained from Monte Carlo simulations. Shades: estimated uncertainty with 1000 samples. SLAEN shows an error-probability scaling advantage over classical classifier.}
    \label{fig:scaling}
\end{figure}

To investigate the performance of SLAEN and the classical classifier with respect to the signal-to-noise ratio of the data, the error probabilities, under the optimum settings for the VQC and the classical SVMs in the 2D data-classification problem, are measured as the margin $\epsilon$ varies in $\{0.2, 0.4, 0.6, 0.8, 1\}$. The results plotted in Fig.~\ref{fig:scaling} show that SLAEN enjoys an error-probability scaling advantage over that of the classical classifier, as manifested in the disparity between the slopes for the two error-probability curves. At $\epsilon = 1$, SLAEN's error probability is more than three-fold less than that of the classical classifier.

\section{Discussions}
{The SLAEN theory paper~\cite{zhuang2019physical} reported an error-probability advantage achieved by an entangled sensor network over that of a sensor network based on separable squeezed states with the same total number of photons, which is verified by the current SLAEN experiment (see Appendix D for details). However, SLAEN's performance has been primarily benchmarked against classical classifiers that do not use any quantum resources \QT{(see Ref.~\cite{zhang2020distributed} for an in-depth discussion about different types of resources used in distributed quantum sensing)}. Such a choice is motivated by two main considerations. First, the compared classical classifier represents a common configuration for sensing and data processing. Introducing quantum resources yields a performance enhancement over the {existing} classical schemes. {In the SLAEN experiment, the power of the coherent-state portion is orders of magnitude stronger than that of either the squeezed or the entangled light, similar to the case in squeezed-light-enhanced Laser Interferometer Gravitational Wave Observatory (LIGO)~\cite{tse2019quantum}. In both cases, the squeezed-light power is limited due to experimental capabilities, so it barely affects the total optical power employed in sensing.} As such, like LIGO, we choose to quantify the quantum advantage as the performance gain over the classical system using the same amount of laser power but taking no advantage of any quantum resources. Second, a complete experimental demonstration of supervised learning based on separable squeezed states requires three independent squeezed-light sources, which places significantly more resource overhead than SLAEN's single squeezed-light source. Hence, SLAEN also enjoys a practical advantage over classical classifiers based on separable squeezed states. \QT{It is worth noting that such a practical advantage would be more pronounced when sensors are nearby so that the entanglement distribution loss is low.} }

\QT{Our experiment has implemented an entanglement source trained by supervised learning. The original SLAEN proposal~\cite{zhuang2019physical}, however, also entails reconfigurable measurements. Since homodyne measurements commute with a linear quantum circuit, SLAEN's performance under three homodyne detectors equals that obtained by the variational measurement apparatus considered by Ref.~\cite{zhuang2019physical}. The current SLAEN protocol only leverages Gaussian sources and measurements, but non-Gaussian resources would potentially improve its performance. Indeed, non-Gaussian measurements have been shown to benefit quantum metrology~\cite{jiang2012strategies}, quantum illumination~\cite{zhang2015entanglement}, and entanglement-assisted communication~\cite{hao2021entanglement}. A variational circuit approach for non-Gaussian entanglement generation and measurements would open a promising route to further enhance the performance.}

\section{Conclusions}
In conclusion, we have experimentally demonstrated the SLAEN framework for quantum-enhanced data classification. Our work opens a new route for exploiting NISQ hardware to enhance the performance of real-world data-processing tasks. Our current experiment verified SLAEN's quantum advantage in classifying features embedded in RF signals, but SLAEN by itself is a general framework applicable to data-processing problems in other physical domains by appropriately engineering entangled probe states and quantum transducers. The present experiment only demonstrated data classification with linear hyperplanes. To accommodate nonlinear hyperplanes, non-Gaussian entangled probe states~\cite{ourjoumtsev2007generation} and joint quantum measurements~\cite{Zhuang2017} would be needed, and the VQC parameter optimizer would also need to be trained to conduct an effective mapping from the SVM hyperplane to the VQC parameters. With these developments, we envisage that SLAEN would create new near-term opportunities in a variety of realms including distributed big-data processing, navigation, chemical sensing, and biological imaging.

\section*{Acknowledgments}
We gratefully acknowledge funding support by the Office of Naval Research Grant No.~N00014-19-1-2190 and the National Science Foundation Grant No.~ECCS-1920742, No.~CCF-1907918, and No.~OIA-2040575. QZ also acknowledges support from Defense Advanced Research Projects Agency (DARPA) under Young Faculty Award (YFA) Grant No.~N660012014029. The authors thank Saikat Guha and William Clark for helpful discussions.
\appendix

\section{OPTIMIZATION ALGORITHM}

\label{sec:algorithm}

The simultaneous perturbation stochastic approximation (SPSA) algorithm is used by the classical support-vector machine (SVM) to update the hyperplane in each training step. The SPSA algorithm calculates an approximation of the gradient with only two measurements, $ {\bm {w_+}}$ and ${\bm {w_-}}$, of the loss function. This simplicity leads to a significant complexity reduction in the cost optimization. See Algorithm~\ref{alg:SPSA} for details.

In the algorithm, $d$ is the dimension of the data set. $d = 2$ for classification problems in a 2D data space, while $d = 3$ for classification problems in a 3D data space. The choices of $a$, $c$, $A$, and $\gamma$ determine the gain sequences $a_k$ and $c_k$, which in turn set the learning rates and have a significant impact on the performance of the SPSA algorithm. The parameters used by the classical SVM in our experiment are: $a=1,c=1,A=200,\alpha=0.602$, and $\gamma=0.1$. 

The SPSA algorithm calls a loss function that is in line with the form of the cost function (Eq. (1) of the main text) but allows for an iterative optimization, as defined below:
\begin{equation}
    {\rm loss}(\bm w, b) =  \left\vert 1-y^{(n)} \left(\bm w \cdot \tilde{\bm \alpha}^{(n)}+b\right)\right\vert_++\lambda \|\bm w \|^2,
\end{equation}

\begin{algorithm}[]
\SetAlgoLined
 Initialization \:
 $a$;$c$;$A$;$\alpha$;$\gamma$;$d$;$N$; $\bm w^{(0)}$; $b^{(0)}$\\
 \For{n=1:N}{
    $a_n=a/(n+A)^{\alpha}$\\
    $c_n=c/n^{\gamma}$\\
    ${\bm {\Delta}_w}=2*{\rm round}({\rm rand}(d,1))-1$\\
    $\bm w_+ = \bm w^{(n-1)}+c_n*{\bm {\Delta}_w}$\\
    $\bm w_- = \bm w^{(n-1)}-c_n*{\bm {\Delta}_w}$\\
    ${\bm {\Delta}_b}=2*{\rm round}({\rm rand}(1,1))-1$\\
    $b_+ = b^{(n-1)}+c_n*{\bm {\Delta}_b}$\\
    $b_- = b^{(n-1)}-c_n*{\bm {\Delta}_b}$\\
    $l_{+}={\rm loss}(\bm w_+,b_+)$\\
    $l_{-}={\rm loss}(\bm w_-,b_-)$\\
    $g_w=(l_{+}-l_{-})/(2*c_n*{\bm {\Delta}_w})$\\
    ${\bm w}^{(n)} = {\bm w}^{(n-1)}-a_n*g_w$\\
    $g_b=(l_{+}-l_{-})/(2*c_n*{\bm {\Delta}_b})$\\
    $b^{(n)} = b^{(n-1)}-a_n*g_b$\\
    }
 \caption{The simultaneous perturbation stochastic approximation (SPSA)~\cite{spall1998overview}
 \label{alg:SPSA}
 }
\end{algorithm}

\section{EXPERIMENTAL DETAILS}
\label{sec:experiment}

\subsection{Experimental setup}
\label{subsec:setup}

\begin{figure*}[bth]
    \centering
    \includegraphics[width=1\textwidth]{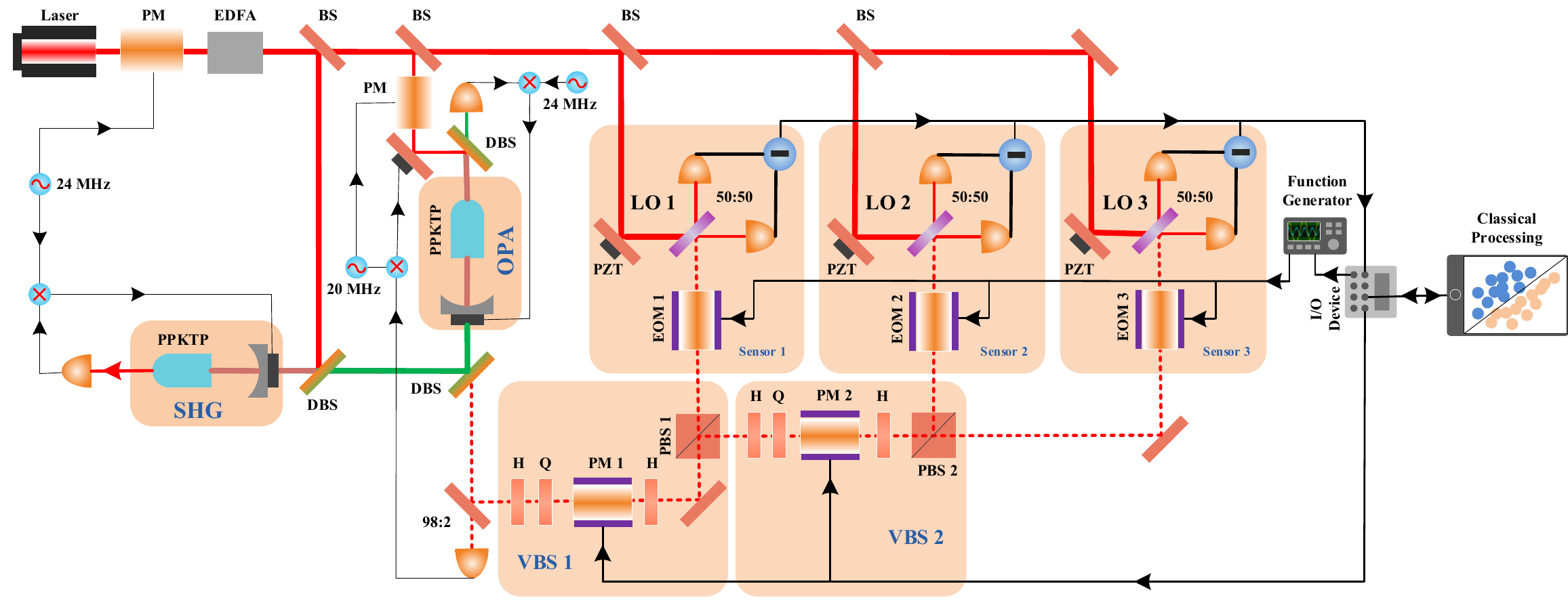}
    \caption{Detailed experiment diagram. EOM: electro-optic modulator; BS: beam splitter; LO: local oscillator; DBS: Dichroic beam splitter; PPKTP: periodically-poled KTiOPO$_4$; OPA: optical parametric amplifier; SHG: secondharmonic generation; PBS: polarizing beam splitter; PM: phase modulator; H: half-wave plate; Q: quarter-wave plate; PZT: piezoelectric transducer; EDFA: erbium-doped fiber amplifier; VBS: variable beam splitter; VQC: variational quantum circuit.}
    \label{fig:exp_detail}
\end{figure*}
A detailed experimental setup is shown in Fig.~\ref{fig:exp_detail}. Squeezed light at 1550 nm is generated from an optical parametric amplifier (OPA) cavity where a type-0 periodically-poled KTiOPO$_4$ (PPKTP) crystal is pumped by light at 775 nm produced from a second harmonic generation (SHG) cavity. The cavities are locked by the Pound-Drever-Hall technique using 24-MHz sidebands created by phase modulating the 1550-nm pump light prior to the SHG. A small portion of light at 1550 nm modulated at 20 MHz is injected into the OPA cavity and phase locked to the pump light to operate in a parametric amplification regime. In doing so, the squeezed light emitted from the OPA cavity is composed of an effective single-mode squeezed vacuum state residing in the 11-MHz sidebands and a displaced phase squeezed state at the central spectral mode. Due to the large quadrature displacement at the central spectral mode, it can be well approximated by a classical coherent state. More details about the characterization of our squeezed-light source are enclosed in Supplemental Material of Ref.~\cite{xia2020demonstration}. 

The squeezed light is directed to a variational quantum circuit (VQC) composed of two variable beam splitters (VBSs) and three phase shifters, parameterized by  $\boldsymbol{v} \equiv \{v_1, v_2, v_3, \phi_1, \phi_2, \phi_3\}$. Here, $v_m$ is the portion of the power diverted to the $m$th sensor, satisfying $\sum_{m=1}^3 v_m=1$. $\phi_m$ is the phase shift on the quantum state at the $m$th sensor.  Each VBS comprises a first half waveplate, a quarter waveplate, a phase modulator (PM), a second half waveplate, and a polarizing beam splitter. The power splitting ratio is controlled by applying a DC voltage generated from a computer-controlled data acquisition board (NI PCI 6115). The DC voltage is further amplified by a high-voltage amplifier (Thorlabs HVA 200) with a gain of $20$ prior to being applied on the PM. The power portions are determined by:
\begin{align}\label{eq:powersplit}
    &v_1=\frac{1}{2}\left(\sin\left(\frac{E_{s_1}}{V_\pi}\pi\right)+1\right) \notag
    \\
    &v_2=1-v_1-v_3 
    \\
    &v_3=\frac{1}{2}\left(\sin\left(\frac{E_{s_2}}{V_\pi}\pi\right)+1\right)(1-v_1) \notag
    \end{align}
where $E_{s_1}, E_{s_2}$ are DC voltages applied on PM 1 and PM 2. 

After the VBSs, the three-mode entangled probe state, represented by the annihilation operators $\{\hat{b}_1,\hat{b}_2,\hat{b}_3\}$, is diverted to three RF-photonic sensors, each equipped with an EOM driven by an RF signal at a 11-MHz carrier frequency. Due to the phase modulation, a small portion of the coherent state at the central spectral mode is transferred to the 11-MHz sidebands, inducing a phase quadrature displacement. The quadrature displacement at each RF-photonic sensor is equal to~\cite{xia2020demonstration}
\begin{eqnarray}
\label{eq:displacement}
   \braket{\hat{b}_m}= \alpha_m &\simeq& i\sqrt{2}\pi g_m a_{c_m} \frac{\gamma E_m}{2V_\pi}  \sin\left(\varphi_m\right),
\end{eqnarray}
where $g_m=\pm 1$ is set by an RF signal delay that controls the sign of the displacement. Choosing $g_m=-1$ is equivalent to introducing a $\pi$-phase shift on the quantum state at the $m$th sensor~\cite{xia2020demonstration}, i.e., setting $\phi_m = \pi$ in the VQC parameters. In Eq.~\eqref{eq:displacement}, $a_{c_m}$ is the amplitude of the baseband coherent state at the $m$th sensor. Specifically, $a_{c_m} = \sqrt{v_m}\beta$, where $\beta$ is the amplitude of the baseband coherent state at the squeezed-light source. $V_{\pi}$ is the half-wave voltage of the EOM, and $\gamma$ describes the conversion from an external electric field $E_m$ to the internal voltage. A more detailed theoretical model for the setup was presented in Ref.~\cite{xia2020demonstration}.

Subsequently, phase-quadrature displacements carried on the quantum light at the three sensors are measured in three balanced homodyne detectors. At each homodyne detector, the quantum light and the local oscillator (LO) first interfere on a $50/50$ beam splitter with a characterized interference visibility of 97\%, and then detected by two photodiodes, each with a $\sim $88\% quantum efficiency. The difference photocurrent is amplified by an transimpedance amplifier with a gain of $20 \times 10^3$ V/A. The DC component of the output voltage signal locks the phase between the LO and the quantum light. The 11-MHz AC component of the voltage signal is demodulated by an electronic mixer, filtered by a 240-kHz low pass filter, and then amplified by a low-noise voltage preamplifier (Stanford Research Systems SR560). The data are acquired by a multifunction I/O device (NI USB-6363) and further processed by a desktop computer in real time. Summing up the measurement data from the three sensors appropriately by $\sum_m \sqrt{v_m} \exp{(i\phi_m)}\tilde{\alpha}_m$ enables the maximal noise reduction, which is equivalent to the noise of the squeezed quadrature of the single-mode squeezed state $\hat{b}$ at the source~\cite{xia2020demonstration}:
\begin{align}
     \label{eq:mode_sum}    {\rm var}(\hat{b})= & {\rm var}\left(\sum_m^M \sqrt{v_m}\exp\left({i\phi_m}\right)\hat{b}_m\right) \notag
     \\
                        = & \frac{1}{4}\left[\frac{\eta}{\left(\sqrt{N_s}+\sqrt{N_s+1}\right)^2}+(1-\eta)\right],
\end{align}
where $\eta$ is the quantum efficiency at each sensor and $N_s$ is the total photon number of the single-mode squeezed light at the source. In our experiment, $\eta \sim 53\%$ and $N_s \simeq 3.3$. In comparing the summation in Eq.~\eqref{eq:mode_sum} with that in Eq.~(1), it becomes clear that choosing the mapping from the hyperplane parameter ${\bm w}$ to the VQC setting $\bm v$ to be $w_m=\sqrt{v_m}\exp\left({i\phi_m}\right)$ minimizes the measurement noise. In the current setup, we measured a 2.9 (3.2) dB noise reduction with the three- (two-) mode entangled state. The characterization of our sensor networks has been reported in Ref.~\cite{xia2020demonstration}. 

\subsection{Calibration}
\label{subsec:calibration}
\subsubsection{Calibration of variational quantum circuit}
\begin{figure*}[tb]
    \centering
    \includegraphics[width=1\textwidth]{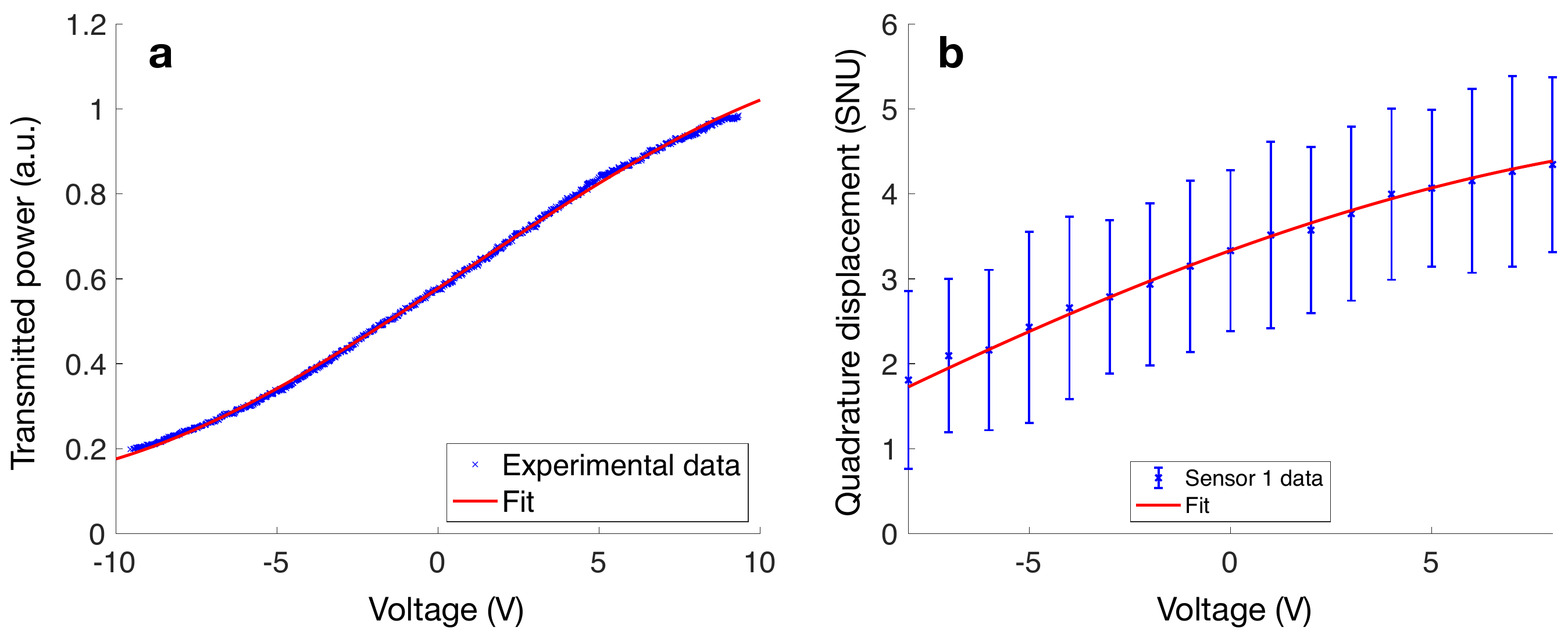}
    \caption{Calibration of the variable beam splitter. (a) Light power delivered to Sensor 1 from VBS 1 (blue crosses) with respect to the input voltage to the high-voltage amplifier, which is proportional to $E_{S_1}$ in Eq.~\ref{eq:powersplit}. Red line: a sinusoidal fit to $v_1$ in Eq.~\eqref{eq:powersplit}, up to a scaling factor. (b) Calibration of quadrature displacement introduced by Sensor 1 vs input voltage to the high-voltage amplifier. Blue crosses: homodyne measurement data at Sensor 1. Red solid line: a fit to $\sqrt{v_1}$ in Eq.~\eqref{eq:powersplit}, up to a scaling factor. Error bars represent the standard deviations of the measurement results, which are determined by the shot-noise level. SNU: shot-noise unit.
    }
    \label{fig:dc_ac_calibration}
\end{figure*}

To ensure accurate configuration of the VQC, we first calibrate the power splitting ratio of both VBSs. In calibrating VBS 1, we scan the voltage $E_{s_1}$ applied on PM 1 and measure the transmitted optical power, as plotted in Fig.~\ref{fig:dc_ac_calibration} (a). The data are fitted to a sinusoidal function in Eq.~\eqref{eq:powersplit}, which derives $V_{\pi} =606$ V for PM 1. An identical calibration procedure is applied on VBS 2, obtain $V_{\pi} = 606$ V for PM 2.

We then measure the quadrature displacements under different VBS splitting ratios, as a means to test the locking stability between the quantum signal and the LO. To do so, while the quantum signal and LO are phase locked, the VBS transmissivity is randomly set to one of 17 values at 30 Hz, subject to the limited bandwidth of the control system. 100 homodyne measurements of quadrature displacement are taken at each transmissivity at a 500 kHz sampling rate. The fitted data are plotted in Fig.~\ref{fig:dc_ac_calibration} (b), showing excellent signal stability and agreement with theory in Eq.~\eqref{eq:powersplit}. The value of the extrapolated $V_{\pi}$ is around 612 V, in good agreement with the specification of the EOM. The tunable range for the VBS transmissivity is between 0.07 and 0.93, limited by the maximum output voltage of the high-voltage amplifier ($\pm 200$ V). VBS 2 is calibrated in an identical way, deriving a $V_{\pi}$ consistent with that of VBS 1. During training, the transmissivity of the VBS is restricted within 0.125 to 0.875 to ensure sufficient light power for phase locking between the quantum signal and the LO.

\subsubsection{Calibration of RF-photonic transduction}
\begin{figure*}[bth!]
    \centering
    \includegraphics[width=1\textwidth]{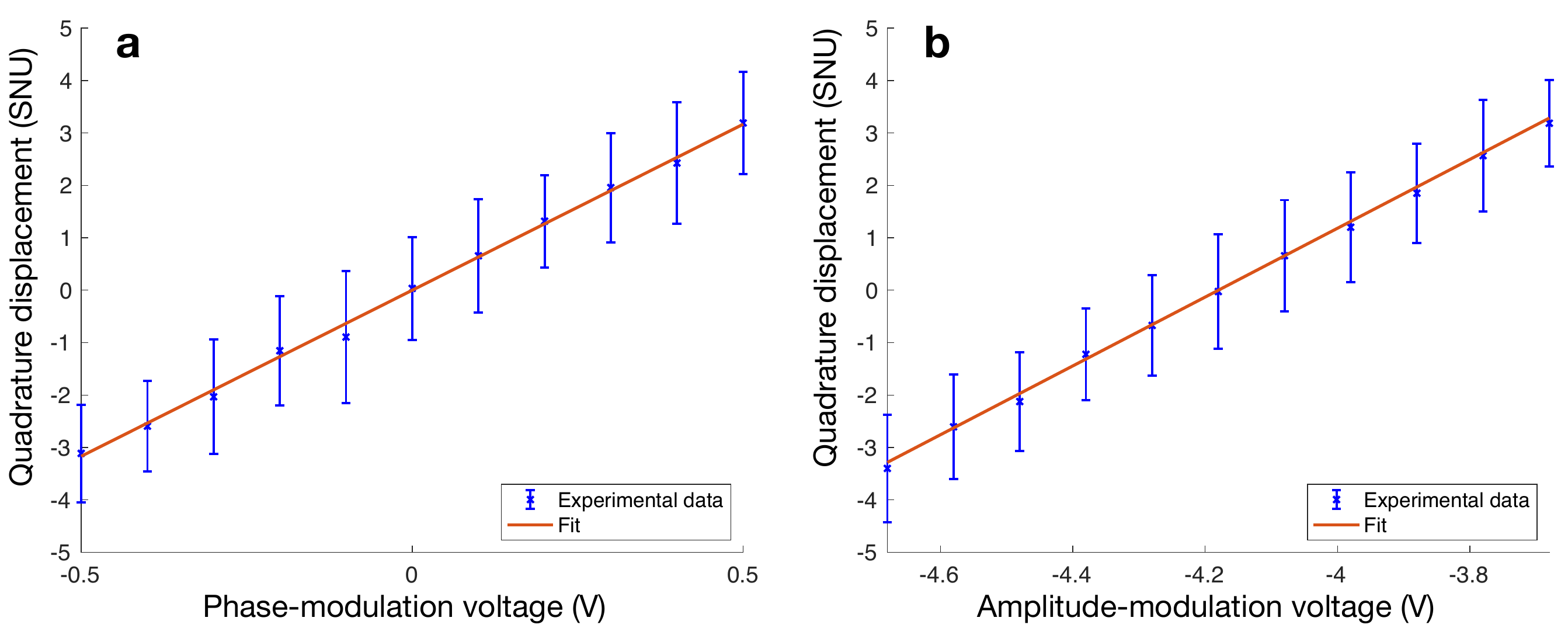}
    \caption{Calibration of RF-photonic transduction. (a) Calibration of the linearity between the quadrature displacement and the phase of the RF signal proportional to the phase-modulation voltage applied on the function generator. (b) Calibration of the linearity between the quadrature displacement and the amplitude of the RF signal proportional to the amplitude-modulation voltage applied on the function generator. Blue crosses: experimental data acquired from homodyne measurements. Red curve: a linear fit. Error bars represent the standard deviations of the measurement results, which are determined by the shot-noise level. SNU: shot-noise unit.
    }
    \label{fig:AM_PM_cali}
\end{figure*}

The training data of RF-field direction (mean-amplitude) classification are prepared by applying phase (amplitude) modulation on the RF-signals. Modulations on RF signals are converted to different quadrature displacements by three EOMs. To ensure linearity in the transduction from the amplitude and phase of the RF signals to quadrature displacements, we calibrate the quadrature displacements at each sensor with respect to the modulation voltages that determine the amplitude and phase for the RF signals applied on the EOMs. In the calibration of phase modulation at Sensor 1 shown in Fig.~\ref{fig:AM_PM_cali} (a), as we sweep the modulation voltage on the function generator for the RF signal from -0.5 V to 0.5 V with an increment of 0.1 V, 100 homodyne measurements of the quadrature displacement are recorded for each modulation voltage at a 500 kHz sampling rate. The distribution of the experimental data on the vertical axis at a given modulation voltage arises from the quantum measurement noise. The fit shows an excellent linear dependence of quadrature displacement vs the modulation voltage. To calibrate the amplitude modulation on the RF signal, we first set the modulation depth to $120 \%$ to allow for a sign flip on RF signal to enable both positive and negative quadrature displacements. We then take 100 homodyne measurements at each modulation voltage at a 500 kHz sampling rate. The experimental data and fit are plotted in Fig.~\ref{fig:AM_PM_cali} (b), showing excellent linear dependence of the measured quadrature displacement with respect to the amplitude of the RF signal. The other two EOMs are calibrated in the same way.

\subsection{Implementation of data classification}
\label{subsec:classification}

\subsubsection{Training stage for SLAEN}

The training stage consists of $N$ steps using randomly produced training data $\left\{\boldsymbol{E}^{(n)},\boldsymbol{\varphi}^{(n)},y^{(n)}\right\}_{n=1}^N$. In the $n$th training data point, $\boldsymbol{E}^{(n)} \equiv \left\{E_m^{(n)}\right\}_{m=1}^M$ and $\boldsymbol{\varphi}^{(n)}\equiv\left\{\varphi_m^{(n)}\right\}_{m=1}^M$ entail, respectively, the probed RF-field amplitudes and phases at the $M=2$ or $M=3$ sensors, and $y^{(n)} \in\{-1,+1\}$ is the true label, which can be derived using the true hyperplane $\left\{\bm w_t, b_t\right\}$ for the data-classification problem in hand. Each sensor then converts the probed RF field into an internal voltage signal, which in turn drives the EOM to induce a quadrature displacement on the quantum signal
\begin{eqnarray}
\label{eq:displacement_training}
   \alpha_m^{(n)} &\simeq& i\sqrt{2}\pi g_m^{(n)} a_{c_m}^{(n)} \frac{\gamma E_m^{(n)}}{2V_\pi}  \sin\left(\varphi_m^{(n)}\right),
\end{eqnarray}
which is similar to Eq.~\eqref{eq:displacement}.

A technicality associated with the quadrature displacement at the $m$th sensor is that it depends on both $E_m^{(n)}$ and the amplitude of the baseband coherent light, $a_{c_m}^{(n)}=\sqrt{v_m^{(n)}}\beta$, as shown by Eq.~\eqref{eq:displacement_training} (see also Ref.~\cite{xia2020demonstration} for a more detailed description). Our experiment focuses on demonstrating the principle of SLAEN, so, without loss of generality, displacement's dependence on the baseband light is eliminated by scaling $\gamma$ by a factor of $1/\sqrt{v_m^{(n)}}$ such that the amount of induced displacement is solely determined by the training data. In our experiment, this is accomplished by applying an extra amplitude modulation that introduces a gain of $1/\sqrt{v_m^{(n)}}$ on the RF signal before it goes to the EOM.

We properly choose $\gamma$ such that the training data point $\{\boldsymbol{E}^{(n)},\boldsymbol{\varphi}^{(n)},y^{(n)}\}$ leads to random quadrature displacements $\bm \alpha^{(n)}$ at the involved sensors with each displacement value initially following a uniform distribution within $[-4,4]$ in the shot-noise unit. The signal-to-noise ratio of the training data set is tuned by excluding points within a margin of $\epsilon$ to the hyperplane. In doing so, the signal-to-noise ratio is raised with an increased $\epsilon$. In the training experiments, $\epsilon=0.6$ is chosen, comparable to one shot-noise unit.

In the 3D data-classification experiment, amplitude modulations on the RF signals from three function generators prepare the training data. Two DC voltages produced by a multifunction I/O device (NI PCI-6115) are used to configure the two VBSs in the VQC. In the 2D data-classification experiment, phase modulations on the RF signals from two function generators prepare the training data, and one DC voltage generated by the same multifunction I/O device is used to configure VBS 1.

\begin{figure}[bth!]
    \centering
    \includegraphics[width=0.45\textwidth]{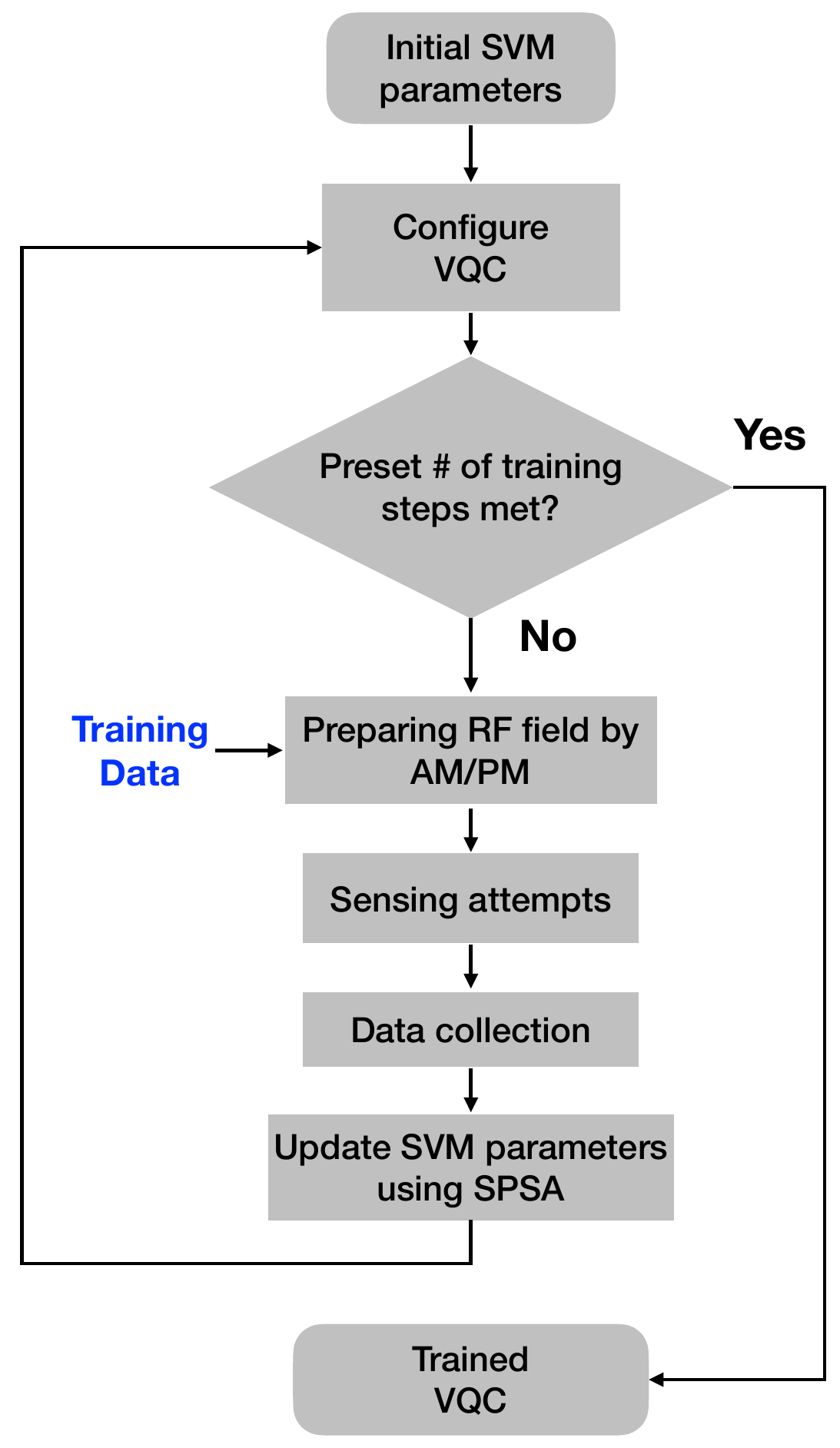}
    \caption{Flowchart of the training process for SLAEN. VQC: variational quantum circuit; SPSA: simultaneous perturbation stochastic approximation. }
    \label{fig:training illustration}
\end{figure}

The flowchart of the training process is sketched in Fig.~\ref{fig:training illustration}. The training starts with an initial hyperplane $\left\{\bm w^{(0)},b^{(0)}\right\}$ and its corresponding VQC setting $\bm v^{(0)}$. Here, $b^{(0)}$ is a number stored in the classical SVM algorithm and will be updated during training. 

The measurement data at each sensor are collected by a multifunction I/O device (NI USB-6363) operating in an on-demand mode and are then transmitted to a desktop computer on which the classical SVM algorithm runs. In the $n$th training step, the measurement data $\tilde{\bm\alpha}^{(n)}$ from all sensors, the true label $y^{(n)}$ and the current hyperplane $\left\{\bm w^{(n-1)},b^{(n-1)}\right\}$ are fed to the SPSA algorithm, which then updates the hyperplane to $\left\{\bm w^{(n)},b^{(n)}\right\}$, as elaborated in Appendix~\ref{sec:algorithm}. The VQC setting is subsequently updated to $\bm v^{(n)}$. The next training step starts with adjusting the power splitting ratios of the VBSs by applying two voltages on the PMs based on Eq.~\eqref{eq:powersplit} and the calibrated $V_\pi$. The new training data are then applied through the EOMs. 

During training, a phase shift $\varphi_m^{(n)}=\pi$ needs to be applied to the quantum state $\hat{b}_m$ when ${\rm sign}(w_m^{(n)})=-1$. Experimentally, this is done by flipping the sign of the emulated RF-signal amplitude. If the sign of the initial hyperplane, ${\rm sign}(w_m^{(0)})$, is different from that of true hyperplane, $w_m^{(n)}$ will move across zero, which will cause zero optical power being delivered to the $m$th sensor such that the phase locking between the quantum signal and the LO breaks down. To avoid this, we restrict the minimum powter splitting ratio to ${\rm min}\left(v_m^{(n)}\right)=0.125$, so that a sign flip on $w^{(n)}_m$ will be applied whenever $v_m^{(n)}$ hits this boundary. The training iterates 200 steps for the 2D data-classification experiment and 390 steps for the 3D data-classification experiment. The loss function converges to its minimum with the hyperplane $\left\{\bm w^\star,b^\star\right\}$ when training completes.

\subsubsection{Utilization stage for SLAEN}

In the utilization stage, SLAEN performs data classification on new measurement data $\tilde{\bm\alpha}^{(n)}$, each with an unknown label. The new data follow the same statistical distribution as the training data. To verify the convergence in the training process, we first measure the error probabilities at different training steps with the hyperplane $\left\{\bm w^{(k)},b^{(k)}\right\}$, where $k\in\{0,20,40,...,160,180,200\}$ in the 2D data-classification experiment and $k\in\{0,30,60,...,330,360,390\}$ in the 3D data-classification experiment. The classical SVM is set to use the hyperplane $\left\{\bm w^{(k)},b^{(k)}\right\}$, and the VQC is configured by the corresponding setting $\bm v^{(k)}$. For each error-probability measurement, 1000 testing data points are applied on the EOMs at a 500-kHz rate by a multifunction I/O device (NI USB-6363), and the measurement data $\tilde{\bm \alpha}^{(n)}$ are synchronously recorded by the same device. A decision is made based on
\begin{equation}
\tilde{y}^{(n)}={\rm sign}\left(\bm w^{(k)} \cdot \tilde{\bm \alpha}^{(n)}+b^{(k)}\right).
\label{eq:label}
\end{equation}
We then estimate the error probability via $P_E=\sum_{n=1}^{N=1000} |(\tilde{y}^{(n)}-y^{(n)})|/N$.

To verify the scaling of error probability with respect to the signal-to-noise ratio of the data set, the hyperplane parameters and the VQC setting are configured to $\left\{\bm w^\star, b^\star\right\}$ and $\bm v^\star$. The error probabilities for the 2D example are measured using data sets with margins $\epsilon\in\{0.2, 0.4, 0.6, 0.8, 1\}$.\\

\subsubsection{Training and utilization stages for classical classifier}
The measurement noise at different sensors in the classical classifier is independent. As such, the classical classifier can solely be trained in post processing carried out by the classical SVM. To perform a direct performance comparison, the training data sets for SLAEN are used to train the classical classifier. The hyperplane $\left\{\bm w^{(n)},b^{(n)}\right\}$ is updated in each training step. The error probabilities at different training steps are measured to validate the convergence. As a comparison, the scaling of the error probabilities for the classical classifier in the 2D example is also measured using the same testing data sets as SLAEN employs.

\subsection{Experiment for general 3D data classification}
\label{subsec:data}

\begin{figure*}[bth!]
    \centering
    \includegraphics[width=1\textwidth]{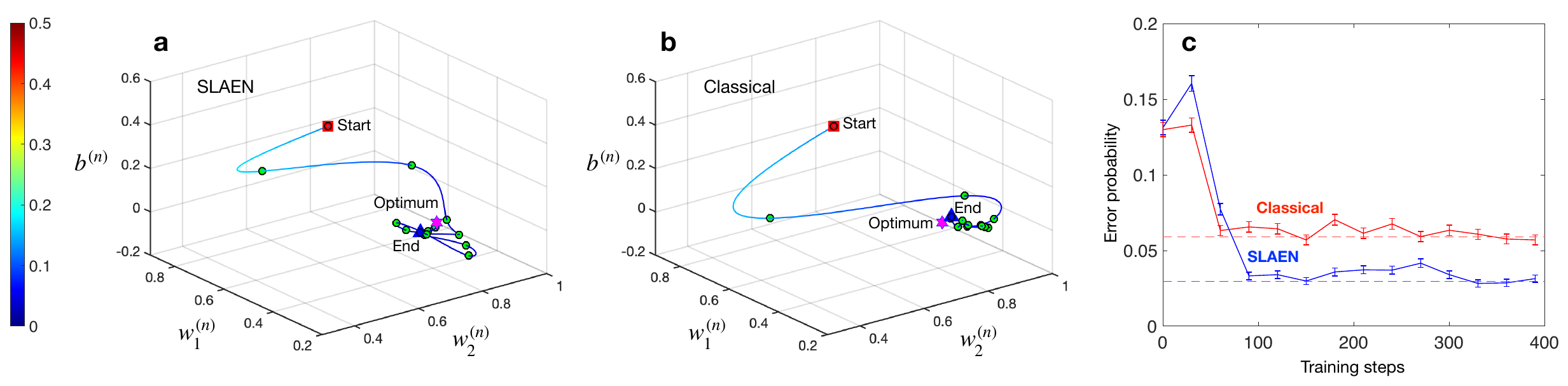}
    \caption{Experimental optimization of hyperplane parameters for a general data-classification problem. Trajectories of hyperplane parameters for SLAEN (a) and for classical classifier (b). Red squares: initial hyperplane parameters prior to training; blue triangles: hyperplane parameters after training; magenta hexagrams: optimum hyperplane parameters. Color gradients: evolution of the error probabilities during training. Green circles: samples of hyperplane parameters at every 30 training steps. Curves are obtained by a cubic spline data fitting. (c) Error probabilities derived under the hyperplane parameters during training. Blue curve:  error probabilities for SLAEN; Blue curve: error probabilities for classical classifier.}
    \label{fig:training_arb}
\end{figure*}

\begin{figure*}[bth!]
    \centering
    \includegraphics[width=1\textwidth]{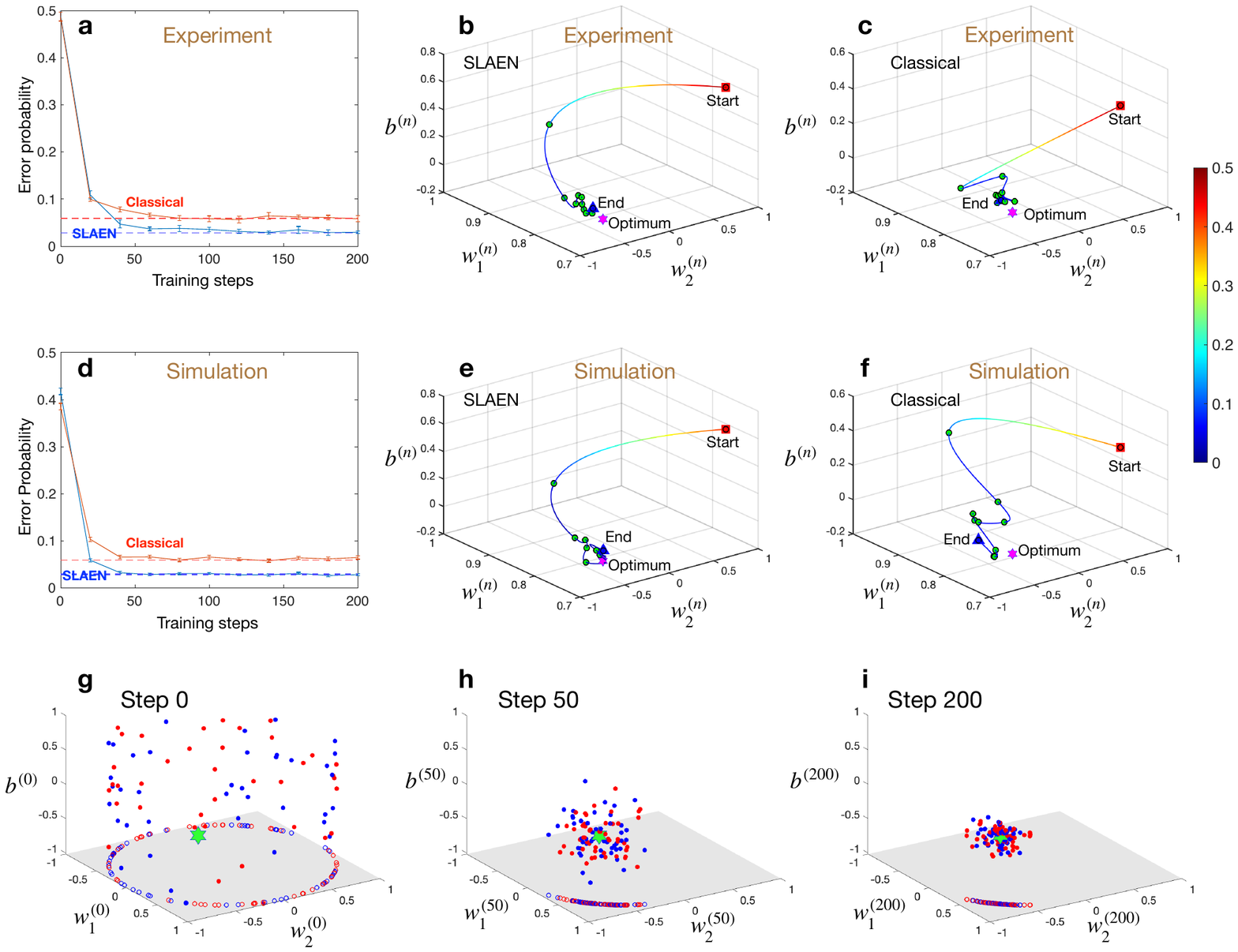}
    \caption{Comparison between experimental data and simulations in training for 2D data classification. (a, d) convergence of the error probabilities during training. Red curve: classical classifier; Blue curve: SLAEN. Horizontal dashed lines: expected error probabilities based on true hyperplanes and measurement-noise levels. Error bars represent one standard deviation of the uncertainty derived from five measurements or simulations each with 1000 data points. (b, e) history of hyperplane parameters for SLAEN during training. (d, f) history of hyperplane parameters for classical classifier. Red squares: initial hyperplane parameters prior to training; blue triangles: hyperplane parameters after training; magenta hexagrams: true hyperplane parameters, representing the optimum. Color gradients: evolution of the error probabilities during training. Green circles: samples of hyperplane parameters at every 20 training steps. Curves obtained by cubic spline data fitting. Simulated distribution of hyperplane parameters prior to training (g), at Step 50 (h), and at Step 200 (i). Blue filled circles: SLAEN hyperplanes; red filled circles: classical-classifier hyperplanes; hexagrams: optimum hyperplanes. Open circles: projected hyperplane parameters onto the $(w_1,w_2)$ face drawn in grey. SLAEN's optimized hyperplanes distribute statistically closer to the optimum solutions.}
    \label{fig:training_sim_2m}
\end{figure*}

To show that SLAEN can be trained to tackle general data-classification problems, we randomly choose a true hyperplane and experimentally train SLAEN and the classical classifier to the undertake the classification task. In the experiment, the initial hyperplane is randomly set to $\left\{\bm w_0 = (0.60, 0.566, 0.566), b_0 = 0.45\right\}$, and the picked true hyperplane is $\left\{\bm w_t = (0.8165, 0.4082, 0.4082), b_t =0 \right\}$. A training data point is supplied to SLAEN at each of the 390 steps, during which the evolving hyperplane parameters are recorded. As anticipated, the experimental result depicted in Fig.~\ref{fig:training_arb} (a) shows that the hyperplane parameters move toward the optimum during training, indicating SLAEN's capability of solving general data-classification problems as long as training data are provided. As a comparison, we train the classical classifier over 390 steps with the same training data set used for SLAEN. The evolving hyperplane parameters during training is plotted in Fig.~\ref{fig:training_arb} (b), showing that the classical classifier can also shift the hyperplane toward the optimum.

With the experimentally measured hyperplane parameters during training, the error probabilities for SLAEN and the classical classifier are derived and plotted in Fig.~\ref{fig:training_arb} (c). SLAEN possesses a clear error-probability advantage over the classical classifier. Specifically, the error probability of SLAEN is two-fold less than that of the classical classifier when both are trained.

\section{SIMULATIONS}
\label{sec:simulations}
We have performed Monte Carlo simulations for the training processes of SLAEN and the classical classifier on a classical computer, as a means to verify the qualitative behaviors of the evolving hyperplane parameters and error probabilities during the training experiments. Note that such a training simulation is merely a testing tool and cannot replace the physical training of SLAEN or the classical classifier in their practical applications because the original data $\{\boldsymbol{E}^{(n)},\boldsymbol{\varphi}^{(n)}\}$ probed by the sensors are in general unavailable.

\subsection{Simulation for two-dimensional data classification}
\label{subsec:2D_sim}
The simulation of the training for 2D data classification undergoes 200 steps, each of which consumes a randomly generated data point. The measurement noise for SLAEN and the classifier is also randomly generated, with the correlation between the measurement noise at SLAEN's different sensors accounted for. To facilitate the comparison between the experimental data and the simulation results, Fig.~3 (a--c) in the main text is replicated as Fig.~\ref{fig:training_sim_2m} (a--c) here. Fig.~\ref{fig:training_sim_2m} (d) depicts the simulated convergence of error probabilities for SLAEN (blue) and the classical classifier (red). In addition, we simulate the evolving hyperplane $\{w_1^{(n)}, w_2^{(n)}, b^{(n)}\}$ during training and plot the results for SLAEN in Fig.~\ref{fig:training_sim_2m} (e) and for the classical classifier in Fig.~\ref{fig:training_sim_2m} (f). In comparing the top and middle panels of Fig.~\ref{fig:training_sim_2m}, excellent qualitative agreement between the experimental data and simulation results is found. Note that since the experimental and simulated measurement results are random, and the SPSA algorithm is stochastic, we only expect a qualitative agreement.

\begin{figure}[bth!]
    \centering
    \includegraphics[width=0.45\textwidth]{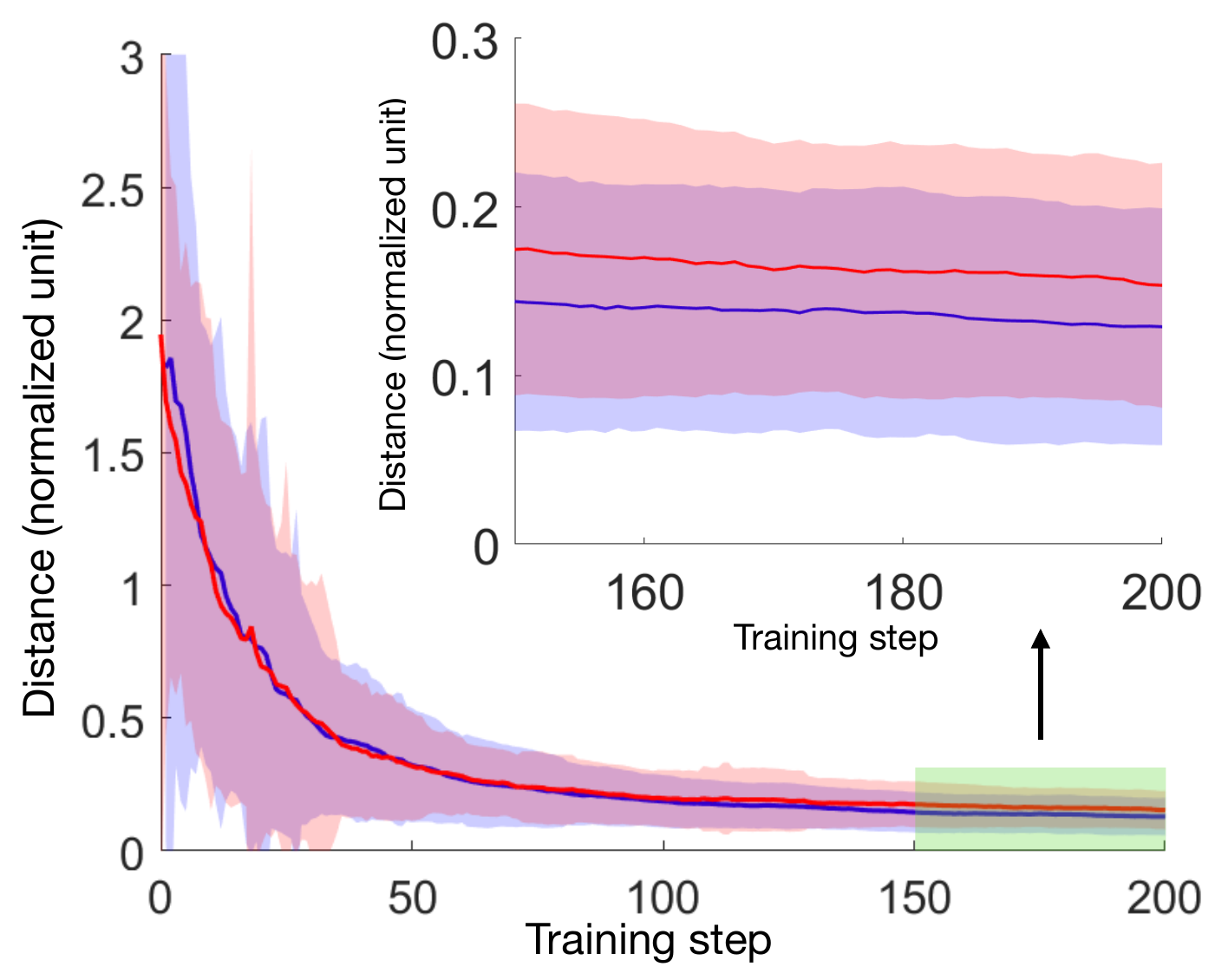}
    \caption{Distance between the true hyperplane and the optimized hyperplanes at different steps under training for 2D data classification. Solid lines: distance averaged over 200 trajectories for the hyperplanes parameters each with a randomly generated initial hyperplane; shaded area: standard deviation of the distances of 200 trajectories. Inset: zoom-in view of the distances between training Step 150 and 200. Blue: SLAEN; red: classical classifier.}
    \label{fig:distance_2m}
\end{figure}

The simulation, in analogy to the experiment, shows that SLAEN's optimized hyperplane (blue triangle) resides closer to the optimum hyperplane (hexagram) than the classical classifier's optimized hyperplane. To investigate whether this is a universal feature for SLAEN, we performed 200 training simulations for both SLAEN and the classical classifier. The initial hyperplane for each training simulation is randomly drawn and is defined as random variables $\{\bm W_S^{(0)}, B_S^{(0)}\}$ for SLAEN and $\{\bm W_C^{(0)}, B_C^{(0)}\}$ for the classical classifier. Fig.~\ref{fig:training_sim_2m} (g) plots the distributions for 50 initial hyperplane parameters in filled circles for both SLAEN (blue) and the classical classifier (red) prior to training. The optimum hyperplane $\{\bm w_t, b_t\}$ is represented by the hexagram. The open circles are the projected hyperplane parameters onto the $(w_1,w_2)$ plane in grey. After 50 training steps, the distributions of the hyperplane $\{\bm W_S^{(50)}, B_S^{(50)}\}$ and $\{\bm W_C^{(50)}, B_C^{(50)}\}$ are drawn in Fig.~\ref{fig:training_sim_2m} (h), showing that the hyperplane parameters are migrating toward the optimum. The distributions of the hyperplane parameters after 200 training steps are depicted in Fig.~\ref{fig:training_sim_2m} (i), which shows, qualitatively, that SLAEN's optimized hyperplanes (blue circles) are almost enclosed by the classical classifier's optimized hyperplanes (red circles). This is an evidence for SLAEN's enhanced accuracy in seeking the optimum solutions.

\begin{figure*}[bth!]
    \centering
    \includegraphics[width=1\textwidth]{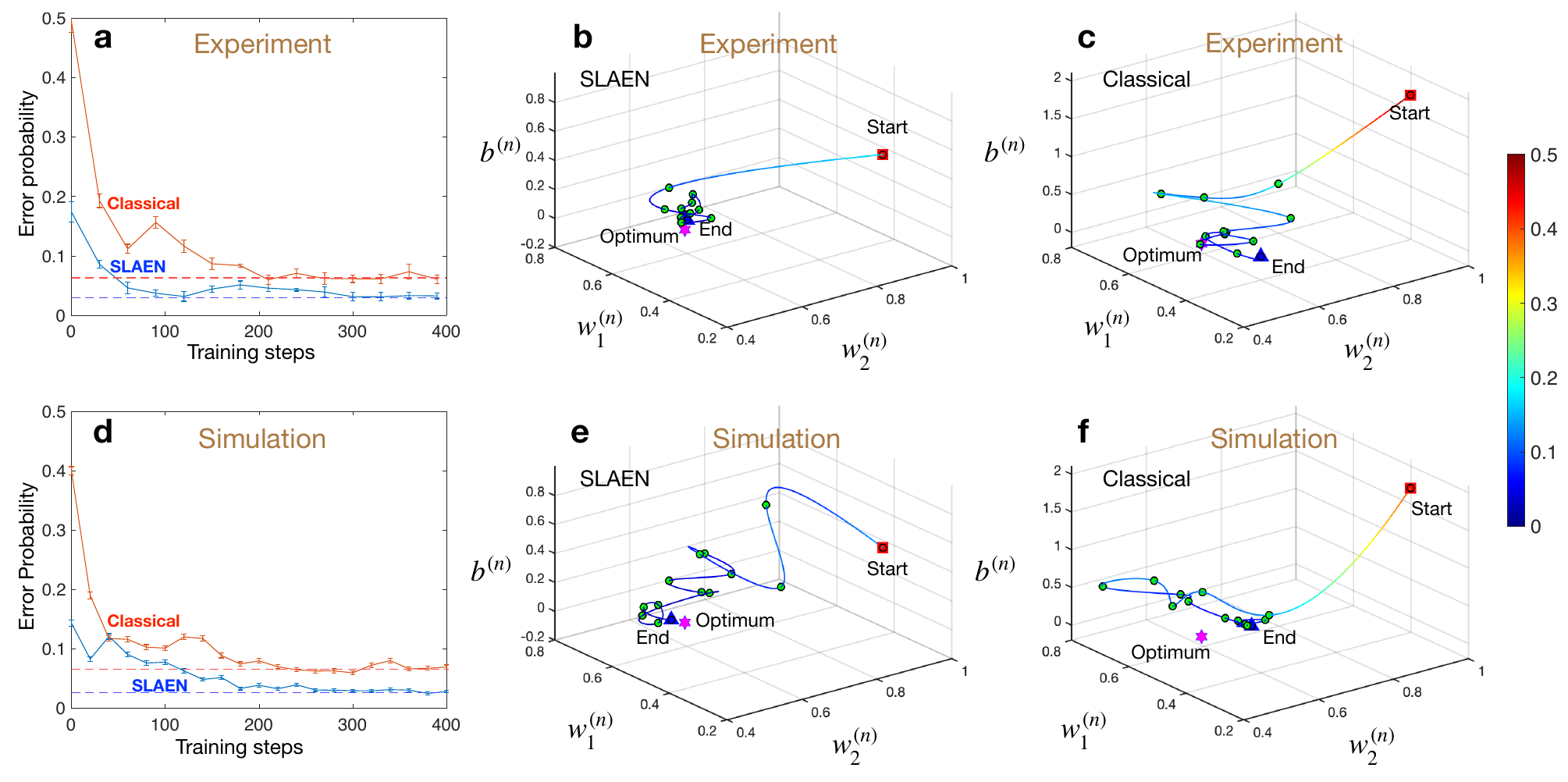}
    \caption{Comparison between experimental data and simulation results of training for 3D data classification. (a, d) convergence of the error probabilities during training. Red curve: classical classifier; Blue curve: SLAEN. Horizontal dashed lines: expected error probabilities based on true hyperplanes and measurement-noise levels. Error bars represent one standard deviation of the uncertainty derived from five measurements or simulations each with 1000 data points. (b, e) history of hyperplane parameters for SLAEN during training. (d, f) history of hyperplane parameters for classical classifier during training. Red squares: initial hyperplane parameters prior to training; blue triangles: hyperplane parameters after training; magenta hexagrams: optimum hyperplane parameters. Color gradients: evolution of the error probabilities during training. Green circles: samples of hyperplane parameters at every 30 training steps. Curves obtained by cubic spline data fitting.}
    \label{fig:training_sim_3m}
\end{figure*}

To conduct a more quantitative assessment on the convergent behaviors for the hyperplane parameters, we define the distance between SLAEN's hyperplanes and the optimum hyperplane after $n$ training steps as
\begin{equation}
    d_S^{(n)} \equiv \left<\sqrt{\left(\bm W^{(n)}_S - \bm w_t\right)^2 + \left(B_S^{(n)} - b_t\right)^2}\right>.
\end{equation}
The standard deviation of the distance is then defined as
\begin{equation}
    \Delta d_S^{(n)} \equiv \sqrt{\left<\left[\sqrt{\left(\bm W^{(n)}_S - \bm w_t\right)^2 + \left(B_S^{(n)} - b_t\right)^2} - d_S^{(n)}\right]^2\right>}.
\end{equation}
Likewise, the distance between the classical classifier's hyperplanes and the optimum hyperplane is defined as
\begin{equation}
    d_C^{(n)} \equiv \left<\sqrt{\left(\bm W^{(n)}_C - \bm w_t\right)^2 + \left(B_C^{(n)} - b_t\right)^2}\right>.
\end{equation}
The standard deviation for the classical classifier's distance is then defined as
\begin{equation}
    \Delta d_C^{(n)} \equiv \sqrt{\left<\left[\sqrt{\left(\bm W^{(n)}_C - \bm w_t\right)^2 + \left(B_C^{(n)} - b_t\right)^2} - d_C^{(n)}\right]^2\right>}.
\end{equation}

The distances at different training steps are plotted in Fig.~\ref{fig:distance_2m} for SLAEN's hyperplanes (red) and the classical classifier's hyperplanes (blue). The distance for SLAEN's hyperplanes after 200 training steps is $d^{(200)}_S = 0.129 \pm 0.07$, as compared to the classical classifier's $d^{(200)}_C = 0.154 \pm 0.073$. The disparity between the distances at the end of training is highlighted via a zoom-in view between Step 150 and Step 200 in the inset of Fig.~\ref{fig:distance_2m}.

\subsection{Simulation for three-dimensional data classification}
\label{subsec:3D_sim}
\begin{figure}[bth]
    \centering
    \includegraphics[width=0.45\textwidth]{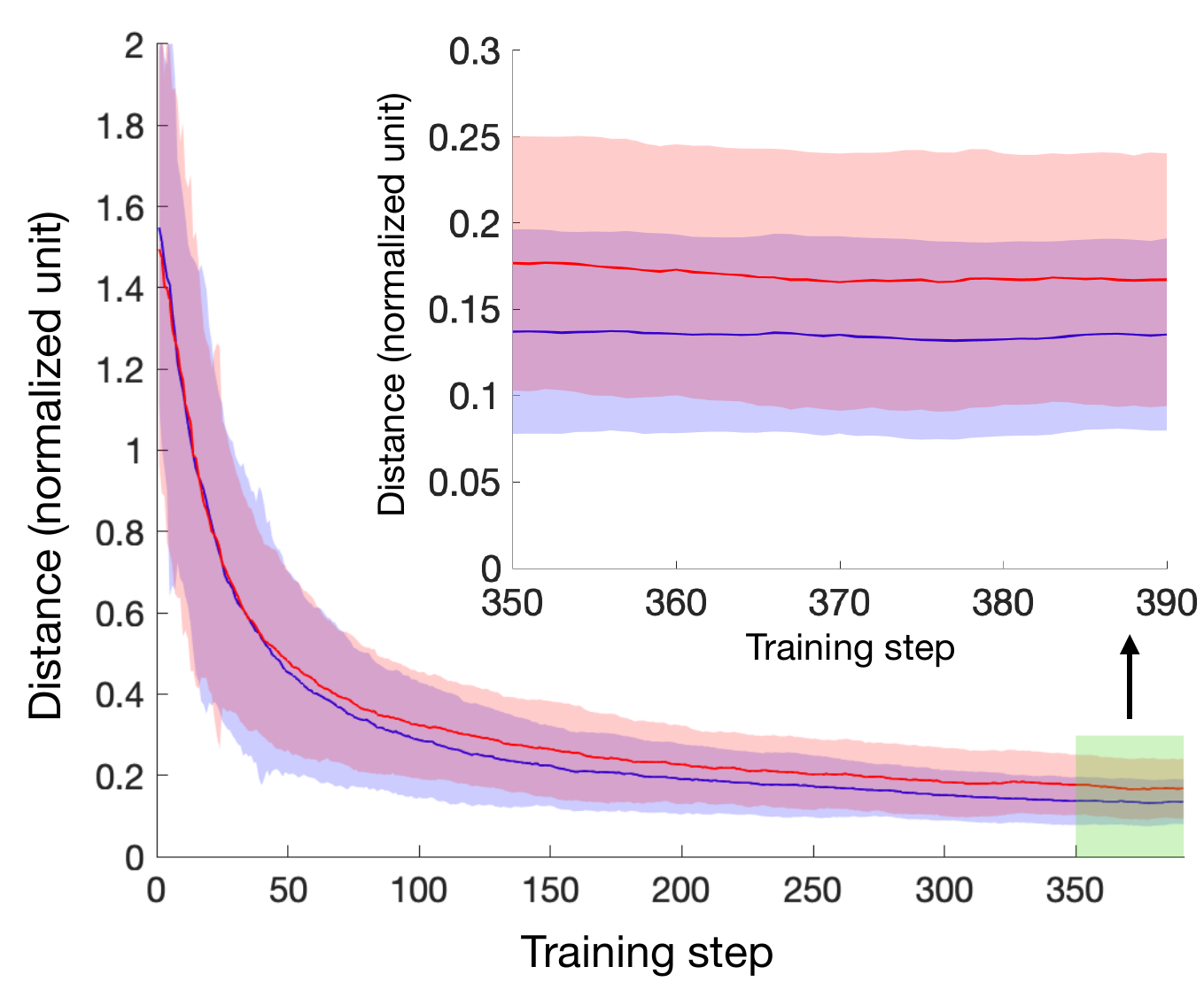}
    \caption{Distance between the true hyperplane and the optimized hyperplanes at different steps under training for 3D data classification. Solid lines: distance averaged over 200 trajectories for the hyperplanes parameters each with a randomly generated initial hyperplane; shaded area: standard deviation of the distances of 200 trajectories. Inset: zoom-in view of the distances between training Step 350 and 390. Blue: SLAEN; red: classical classifier.}
    \label{fig:distance_3m}
\end{figure}

We next simulate the training processes of SLAEN and the classical classifier for 3D data classification. The training for each case takes 390 steps, identical to the number of training steps in the experiment. To facilitate the comparison between experimental data and the simulation results, the plots in Fig.~3 (d--f) of the main text are replicated as Fig.~\ref{fig:training_sim_3m} (a--c) here. The simulated convergence of error probabilities is plotted in Fig.~\ref{fig:training_sim_3m} (d). Fig.~\ref{fig:training_sim_3m} (e) and (f) draw, respectively, the simulated histories of the hyperplane parameters for SLAEN and the classical classifier during training. The qualitative behaviors for the experimental data agree very nicely with those of the simulation results, thereby supporting the validity of the experimental approach.

In addition, we conducted a statistical study of the distances between the hyperplanes and the optimum hyperplane during training for 3D data classification. The distributions of the hyperplane parameters for SLAEN and the classical classifier are plotted in Fig.~3 (g--i) of the main text for, respectively, the initial hyperplanes, the hyperplanes after 100 training steps, and the hyperplanes when training completes. It can be visually observed that SLAEN's optimized hyperplanes locate closer to the optimum hyperplane than the classical classifier's optimized hyperplanes. As a quantitative analysis, the distances vs. training step curves for SLAEN and the classical classifier are plotted in Fig.~\ref{fig:distance_3m}. Akin to Fig.~\ref{fig:distance_2m}, SLAEN enables a reduced distance between its optimized hyperplanes and the optimum hyperplane. This is a consequence of the entanglement-enabled measurement-noise reduction mechanism that SLAEN harnesses. The inset of Fig.~\ref{fig:distance_3m} is a zoom-in view of the distances for SLAEN and the classical classifier near the end of training. After 390 training steps, we define SLAEN's optimized distance as $d_S \equiv d_S^{(390)} = 0.135\pm 0.056$ and the classical classifier's optimized distance as $d_C \equiv d_C^{(390)} = 0.167 \pm 0.073$. Both $d_S$ and $d_C$ are reported in the main text.

\section{DATA CLASSIFICATION USING SEPARABLE SQUEEZED STATES}
The original SLAEN theory paper~\cite{zhuang2019physical} showed that the performance of data processing tasks undertaken by an entangled sensor network is superior to that of a sensor network based on separable squeezed states that have the same total photon number as the entangled state. In this Section, we show, in simulation and by experimental data, that our SLAEN experiment achieves an advantage over a sensor network based on separable squeezed states in data-classification tasks, subject to a photon-number constraint. Specifically, our simulation shows that data classification based on separable squeezed states has a larger error probability than that of our SLAEN experiment. We also experimentally show that the quantum noise of a sensor network with separable squeezed states is higher than that of SLAEN, thereby supporting the SLAEN's claimed advantage over supervised learning based on separable squeezed states. Finally, we present the motivation behind the main text's focus on a performance comparison between SLAEN and classical classifiers based on coherent states.

\subsection{Simulation for 3D data classification using separable squeezed states}
\begin{figure}[h!]
    \centering
    \includegraphics[width=0.45\textwidth]{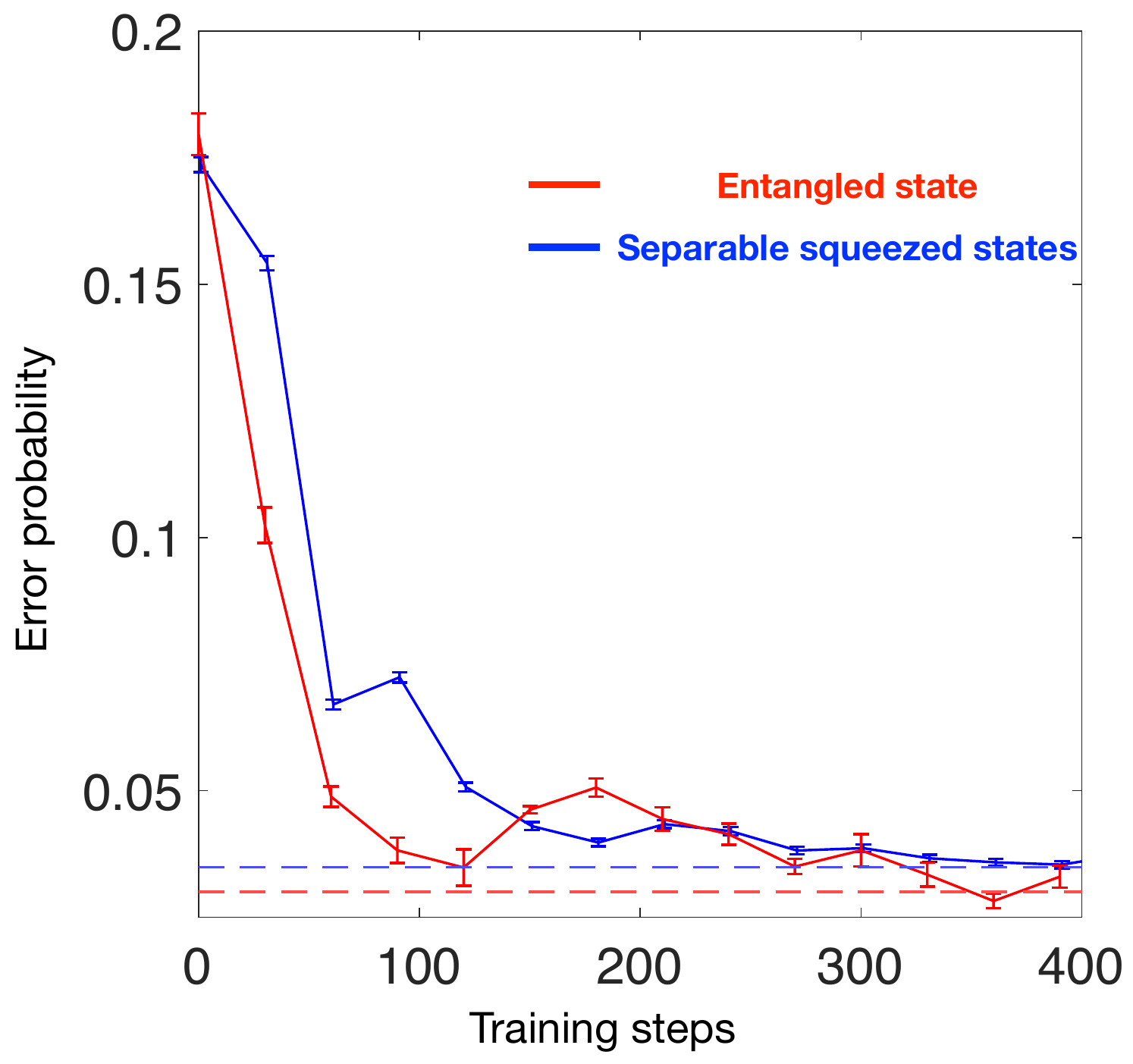}
    \caption{Convergence of error probability. Blue curve: simulation of training process based on separable squeezed states. Error bars represent one standard deviation of the uncertainty derived from 60000 simulated data points. Red curve: SLAEN experiment. Error bars represent one standard deviation of the uncertainty derived from 10 measurements. Horizontal dashed
    lines: expected error probabilities based on true hyperplanes and measurement-noise levels. }
    \label{fig:separable_sim}
\end{figure}

We simulate the training process of average RF-field amplitude classification undertaken by a three-node sensor network based on separable squeezed states. The total photon number of the separable squeezed states is set to be the same as that of the entangled states in the SLAEN experiment. The evolution and convergence of the error probabilities during the training process are plotted in Fig.~\ref{fig:separable_sim}. In the simulation, the initial hyperplane is set to $\left\{\bm w_0 = (0.9044, 0.3152, 0.2876), b_0 = 0.53\right\}$, identical to that of the training in the SLAEN experiment. As a comparison, we experimentally measure the error-probability evolution in SLAEN by taking ten more measurements based on the same experimental setting as what is used to produce Fig.~\ref{fig:training_sim_3m}a. Our SLAEN experiment shows an error probability advantage of $\sim 13\%$ over that of a simulated sensor network based on separable squeezed states.

\subsection{Experimental noise calibrations}
\begin{figure}[h!]
    \centering
    \includegraphics[width=0.45\textwidth]{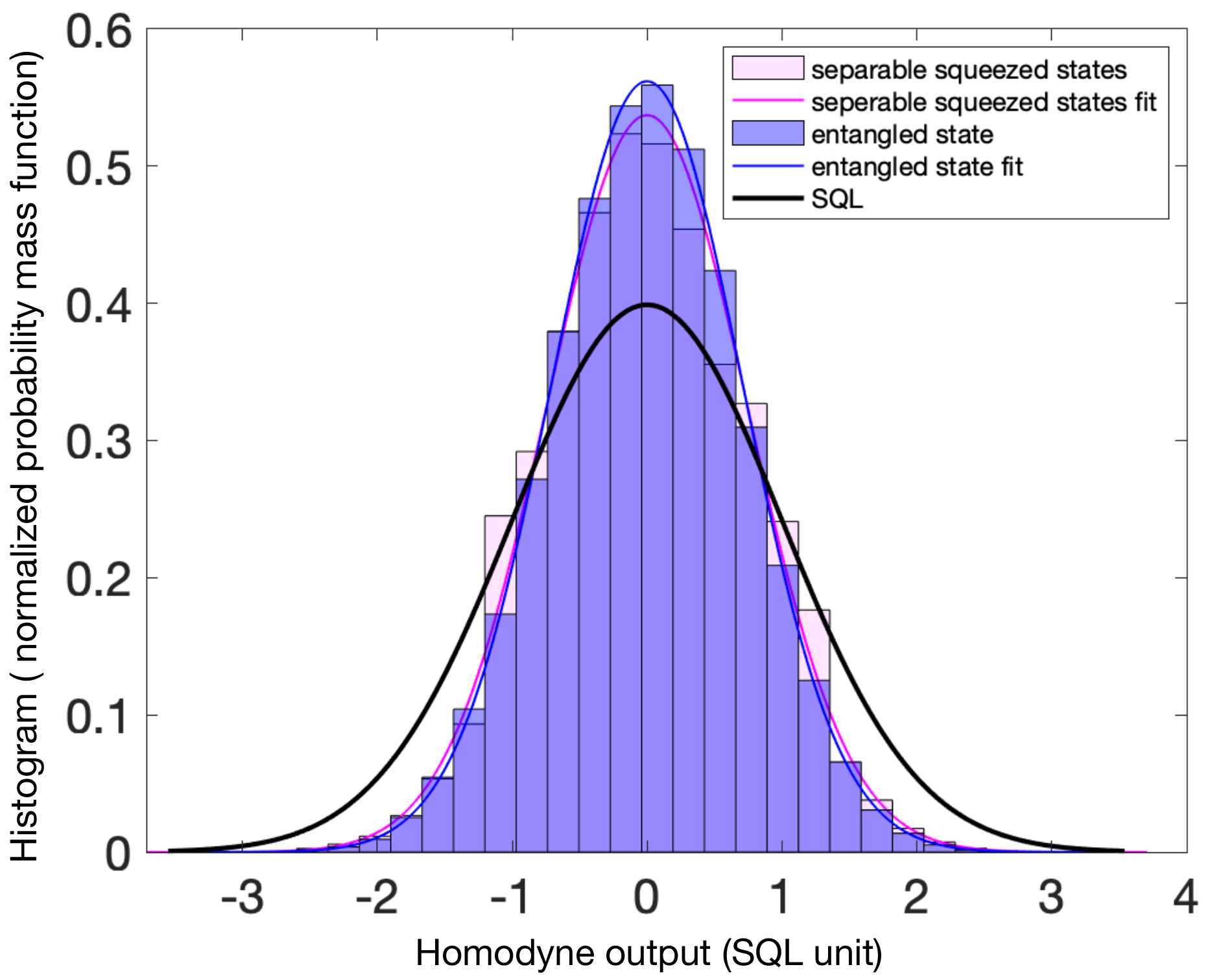}
    \caption{Histograms of homodyne data. Red bins: separable squeezed states. Blue bins: entangled state. Histograms are normalized to probability mass functions and fitted with Gaussian probability density functions. Red curve: theory fit for separable squeezed states. Blue curve: theory fit for entangled state. Black curve: standard quantum limit (SQL). }
    \label{fig:separable_exp}
\end{figure}

A complete demonstration of a three-node sensor network with separable squeezed states requires three independent squeezed-light sources, which places a significant resource overhead. Instead, we calibrate the quantum noise of a sensor network with separable squeezed states using a time-domain multiplexing approach introduced by Ref.~\cite{guo2020distributed}. We first set the mean photon number of our squeezed-light source to that of a separable squeezed state at a single sensor. We then take three samples in the time domain to emulate the independent quantum noise at three sensor nodes. The histogram of the averaged homodyne data is plotted in Fig.~\ref{fig:separable_exp} and fitted with a normalized Gaussian probability density function. Since the measured noise variance of the separable sensor network is $\sim 11.7\%$ higher than that of SLAEN, it is anticipated that the error probability of SLAEN beats that of a sensor network based on separable squeezed states. 

\subsection{Performance comparison}
Quantum metrology studies how nonclassical resources such as squeezed light and entanglement can be utilized in a measurement system to enable a performance advantage over systems based on classical resources. Such a performance gain in sensing underpins SLAEN's error-probability advantage over separable sensor networks. In many practical optical sensing systems such as the Laser-Interferometer Gravitational-Wave Observatory (LIGO), the usable power of the classical laser light is limited due to, e.g., thermal effects, photon radiation-pressure induced torques, and parametric instabilities that cause adverse effects on the system performance~\cite{tse2019quantum}. Nonclassical squeezed light is then injected into the system to further improve the measurement sensitivity. In such a scenario, the measurement sensitivity achieved by a classical laser at a given power level is defined as the standard quantum limit (SQL). Surpassing the SQL using nonclassical resources demonstrates a quantum advantage enabled by quantum metrology. In our experiment, SLAEN's performance is compared with that of a classical classifier based on laser light, i.e., coherent states. The error probabilities for a classical classifier are measured at a given laser power level. SLAEN's error probabilities are then measured at the same laser power level while entanglement is distributed and shared by the sensors. In the SLAEN experiment, the calibration gives a total photon number of the entangled state of $N_S=3.3$ and the quantum efficiency of $\eta=0.53$ at each sensor. In a conceived three-node sensor network based on separable squeezed states, the mean photon number of a separable squeezed state at each sensor will be $1.1$, so that the total photon number matches that of the entangled state. We can then estimate a noise reduction of $2.57$ dB below SQL at the same quantum efficiency at each sensor as the SLAEN experiment. The squeezed state residing at the 11-MHz sidebands is at tens of pico Watts power level, while in the experiment most photons at each sensor originate from the strong ($\sim 50$ $\mu$W) coherent state at the central wavelength of 1550 nm. Give the $\sim 6$ orders of magnitude power disparity between the strong coherent state and the quantum states at the sidebands, a separable sensor network based on either coherent states or separable squeezed states employs nearly identical optical power as an entangled sensor network. This situation is in analogy to LIGO in which the overall optical power remains nearly unchanged despite the injection of squeezed light into the interferometer.
%

\end{document}